\documentclass[aps,prb,amsmath,amssymb,showpacs]{revtex4}
\usepackage{graphicx}
\usepackage{bm}
\usepackage{bbold}

\def\br{ \bm{r} }
\def\bk{ \bm{k} }
\def\bq{ \bm{q} }
\def\bK{ \bm{K} }
\def\bgam{ \bm{\gamma} }
\def\im{ \mathrm{Im}\, }
\def\re{ \mathrm{Re}\, }
\def\sgn{\, \mathrm{sgn}\, }
\def\tr{\,\mathrm{tr}\,}
\def\trb{\,\mathrm{tr}_B}
\def\trn{\,\mathrm{tr}_N}
\def\Tr{\,\mathrm{Tr}\,}

\begin{document}
\title{Symmetry and topology of two-dimensional noncentrosymmetric superconductors}

\author{K. V. Samokhin\footnote{e-mail: kirill.samokhin@brocku.ca}}

\affiliation{Department of Physics, Brock University, St. Catharines, Ontario L2S 3A1, Canada}
\date{\today}

\begin{abstract}
We present a detailed study of the gap symmetry and the quasiparticle wave function topology in two-dimensional superconductors without inversion center. 
The strong spin-orbit coupling of electrons with the crystal lattice makes it necessary to describe superconductivity in terms of one or more nondegenerate bands characterized by helicity. 
We develop a topological classification of the superconducting states using the integer-valued Maurer-Cartan invariants and the Bogoliubov Wilson loops, and also calculate the spectrum of fermionic boundary modes.
\end{abstract}

\pacs{74.20.-z}


\maketitle

\section{Introduction}
\label{sec: Intro}

Superconductors without inversion symmetry have become one of the most studied classes of materials in the past decade, see Ref. \onlinecite{NCSC-book} for a review and references. The long list of noncentrosymmetric superconductors includes compounds 
both with and without strong electron correlations, the former exemplified by the heavy-fermion materials CePt$_3$Si (Ref. \onlinecite{CePt3Si}), Ce(Rh,Ir)Si$_3$ (Ref. \onlinecite{CeRhSi3}), and UIr (Ref. \onlinecite{UIr}), while
the latter is represented by Y$_2$C$_3$ (Ref. \onlinecite{Amano04}), Li$_2$(Pd$_{1-x}$,Pt$_x$)$_3$B (Ref. \onlinecite{LiPt-PdB}), SrPtSi$_3$ (Ref. \onlinecite{SrPtSi3}), and many others. 
These materials have properties that are very different from the predictions of the Bardeen-Cooper-Schrieffer (BCS) theory of superconductivity, due to the qualitative changes in the electron band structure produced by 
the spin-orbit (SO) coupling of electrons with the crystal lattice. 

The SO coupling in a noncentrosymmetric crystal lifts the spin degeneracy of the electron states almost everywhere in the Brillouin zone (BZ), producing nondegenerate Bloch bands labelled by ``helicity'' and endowing 
the quasiparticle wave functions 
with a complex spin structure and a nontrivial momentum-space topology. This is responsible for a number of novel properties, such as the anomalous de Haas-van Alphen and Hall effects in the normal state,\cite{Sam09} 
the magnetoelectric effect,\cite{ME-effect} novel nonuniform superconducting states in the presence of magnetic field\cite{helical-states} or even without any field,\cite{nonuniform-zero-field}
the topologically protected gapless boundary modes and quantum spin Hall effect,\cite{SF09,TYBN09} 
and the unusual impurity response in the superconducting state.\cite{Sam-chapter} If the SO splitting of the bands is large compared to all superconducting energy scales,
then the Cooper pairing occurs only between quasiparticles with the same helicity, with profound consequences for the pairing symmetry.\cite{GR01,SCSam-04} 

Another reason for a recent surge of interest in noncentrosymmetric superconductors is that they appear to be promising candidates for topological superconductivity. The defining property of topological quantum matter is 
that the quantum states in the bulk fall into distinct classes characterized by integer-valued topological invariants, which remain unchanged under sufficiently small variations of the system's parameters.\cite{Volovik-book,top-SC} 
One popular example of a centrosymmetric topological superconductor or superfluid is the time reversal (TR) symmetry-breaking chiral $p$-wave state, which is realized in Sr$_2$RuO$_4$ (Ref. \onlinecite{SrRuO}) and 
thin films of superfluid ${}^3$He-$A$ (Ref. \onlinecite{VY89}). Superfluid ${}^3$He-$B$ is an example of a TR-invariant topological state, which is similar to a three-dimensional (3D) topological insulator.\cite{QHRZ09}
A common feature of the topological materials is that, while fermionic excitations in the bulk are gapped, there are gapless states localized near various inhomogeneities. 
Archetypal examples include the current-carrying boundary modes in the quantum Hall insulators\cite{Hal82} or topological band insulators,\cite{KM05} 
the Andreev bound states near the sample surfaces or the domain walls separating different degenerate ground states,\cite{DW-fermions} and 
the Majorana fermions in the vortex cores in chiral $p$-wave superconductors.\cite{KS91}

Noncentrosymmetric superconductivity has mostly been observed in 3D materials, but it can also be realized in two dimensions (2D). Always a subject of a considerable interest, both experimental and theoretical, 
the field of 2D superconductivity has recently received a big boost, following the discovery of superconductivity in the 2D electron gas 
at the interface LAO/STO between two band insulators, LaAlO$_3$ and SrTiO$_3$ (Ref. \onlinecite{LAO-STO}). Other similar systems include the interfaces LSCO/LCO between metallic and insulating copper oxides, 
or LTO/STO between a Mott insulator LaTiO$_3$ and SrTiO$_3$, and also surfaces of doped insulating oxides, such as STO and possibly WO$_3$, see Ref. \onlinecite{interface-SC} for a review and references. 
A particularly attractive feature of the interface superconductors is that their electronic properties, in particular, the carrier density and the SO coupling strength, can be controlled by an external electric field. 
The superconducting critical temperature $T_c$ in these systems can be as high as 109 K, for FeSe single layers on doped STO substrates.\cite{FeSe-layers} 
The interface superconductors lack inversion symmetry, due to the different nature of the materials sandwiching the conducting layer. Furthermore, since the SO coupling is much larger than $T_c$ (Ref. \onlinecite{SOC-2D}),
the 2D electron bands are well split, suppressing any interband Cooper pairing. 

The standard theoretical approach, which relies on the separation of singlet and triplet pairing 
channels in spin-degenerate bands, works for centrosymmetric superconductors, see, \textit{e.g.}, Ref. \onlinecite{Book}, but is not justified in materials with a large SO splitting of nondegenerate bands.  
In the latter case, it is appropriate to do away with the spin representation and construct the Cooper pairs from the time-reversed helicity band states. 
The band representation is essential for any BCS-like model of pairing built upon the conceptual foundation of the Fermi-liquid theory, in which fermionic quasiparticles exist and experience an attractive
interaction only in the vicinity of the Fermi surface. The Fermi surface itself can only be defined in the band representation, as a set of points in momentum space where a helicity band crosses the Fermi level.   

The goal of this paper is to develop a detailed theory of the pairing symmetry and the topological properties of 2D crystalline noncentrosymmetric 
superconductors in the helicity band representation, which is introduced in Sec. \ref{sec: electron bands}. In contrast to previous studies, our treatment neither uses the spin representation nor relies on any particular model of the SO coupling, 
although the generalized Rashba model is sometimes used to illustrate our results. In Sec. \ref{sec: SC}, starting with a modified BCS model of pairing in nondegenerate bands, we analyze all possible symmetries of the 
superconducting gap and derive the proper form of the Bogoliubov-de Gennes (BdG) Hamiltonian for the fermionic quasiparticles. In Secs. \ref{sec: GFs} and \ref{sec: Berry phases}, 
we develop two different schemes of topological classification of the superconducting states, using the Maurer-Cartan invariants built from the Green's functions and the Wilson loops 
containing the Berry potential of the Bogoliubov quasiparticles, and also calculate the spectrum of the fermionic boundary modes.  
Throughout the paper we use the units in which $\hbar=k_B=1$, neglecting, in particular, the difference between the quasiparticle momentum and wavevector.

\section{Electron bands in two dimensions}
\label{sec: electron bands}

Consider a 2D electron gas confined in the $xy$ plane. We assume that its environment lacks ``upside-down'' symmetry under the reflection $z\to-z$, and therefore the system does not have a center of inversion.
The symmetry group of the normal state includes operations from the space group of the 2D crystal lattice as well as the TR operation $K$. 
There are two contributions to the SO coupling of electrons with the lattice, one originating from the inversion-symmetric part of the lattice potential,
including the atomic cores, which is particularly important in compounds with heavy elements, and the other coming from the inversion-antisymmetric part of the potential, which is sensitive to the spatial arrangement 
of the atoms. While the former contribution merely replaces spin with pseudospin and does not lift the band degeneracy, it is the latter that plays a crucial role in noncentrosymmetric systems.\cite{Sam09}

In the presence of both TR and inversion symmetries, the electron bands are twofold degenerate at each wavevector $\bk$ in the 2D BZ (due to the momentum space periodicity, the BZ is topologically equivalent to a torus). Indeed, the
states $|\bk\rangle$ and $KI|\bk\rangle$, where $I$ is the inversion, correspond to the same $\bk$, are orthogonal, and have the same energy. In addition, these two states are degenerate with another pair of orthogonal
states, $K|\bk\rangle$ and $I|\bk\rangle$, which correspond to $-\bk$. Starting with some $\bk$ in the irreducible part of the
BZ, one can construct a basis of the pseudospin Bloch states at all other wavevectors in the star of $\bk$, which transform under the symmetry operations in the same way as the pure
spin states.\cite{UR85} Below we use the same notation $\alpha,\beta=\uparrow,\downarrow$ for the spin or pseudospin direction. The four degenerate pseudospin states corresponding to $\bk$ and $-\bk$ 
can be used to introduce the singlet and triplet superconducting order parameters, with simple transformation properties under the point group and TR operations.\cite{Book}

In contrast, in noncentrosymmetric crystals with the SO coupling only the TR operation is available to connect the states of the same energy corresponding to $\bk$ and $-\bk$. 
Therefore, the pseudospin degeneracy of the bands is lifted almost everywhere in the 2D BZ, except some high-symmetry points. We shall see below that the band degeneracy points (also called the Weyl points) 
play a crucial role in determining the Bloch states' topology. In the limit of zero SO coupling but still without inversion symmetry, double degeneracy of the bands at all $\bk$ is preserved by arbitrary spin rotations. 
We do not consider this case here, because it is not applicable to real noncentrosymmetric superconductors.

We will use the index $n$, called the helicity, see Sec. \ref{sec: Rashba model} below, to label the nondegenerate electron bands, which contain all information about the lattice potential and the SO coupling, 
so that $|\bk,n\rangle$ is the lattice-periodic part of the Bloch wave function belonging to wavevector $\bk$ in the $n$th band. The corresponding band dispersion, $\xi_n(\bk)$, includes the chemical potential 
(we neglect the difference between the chemical potential and the Fermi energy $\epsilon_F$) and is even in momentum:
\begin{equation}
\label{xi-even}
  \xi_n(\bk)=\xi_n(-\bk).
\end{equation}
This last property is a consequence of TR symmetry, because the states $|\bk,n\rangle$ and $K|\bk,n\rangle$ belong to $\bk$ and $-\bk$, respectively,
and have the same energy. Recall that the TR operator for spin-1/2 particles has the form $K=i\hat\sigma_2K_0$, where $\hat\sigma_2$ is the Pauli matrix and $K_0$ is complex conjugation. 
Since the bands are nondegenerate almost everywhere in the BZ, one can write 
\begin{equation}
\label{t_n}
  K|\bk,n\rangle=t_n(\bk)|-\bk,n\rangle,
\end{equation}
where $t_n(\bk)$ is a phase factor, which is not gauge invariant, in the sense that it depends on the phases of the Bloch states.  Since $K^2|\bk,n\rangle=-|\bk,n\rangle$ and, on the other hand, 
$K^2|\bk,n\rangle=t^*_n(\bk)t_n(-\bk)|\bk,n\rangle$, we obtain: $t_n(-\bk)=-t_n(\bk)$. The phase factors $t_n$ are not defined at $\bk=\bm{0}$ or any other TR invariant point in the BZ. 
A momentum-space point $\bk=\bK$ is called TR invariant if it satisfies the condition $-\bK=\bK+\bm{G}$, where $\bm{G}$ is a reciprocal lattice vector. The total number of the TR invariant points in the BZ is always even and 
denoted by $N_{\mathrm{TRI}}$. At TR invariant momenta, the $n$th band is degenerate with another band, see Sec. \ref{sec: Rashba model} below. 

Since the Cooper pairing takes place between the Bloch states $|\bk,n\rangle$ and $K|\bk,n\rangle$, corresponding to opposite momenta near the Fermi surface, and the superconducting order parameter varies little on the scale of the inverse Fermi 
wavevector, one can neglect the lattice translations and focus on the crystallographic rotations and reflections in the vertical planes, which form the 2D point group $G$. There are ten crystallographic point groups in 2D, 
also known as the rosette groups: five cyclic groups $\mathbf{C}_n$ and five dihedral groups $\mathbf{D}_n$, with $n=1$, $2$, $3$, $4$, or $6$. 
The cyclic group $\mathbf{C}_n$ has $n$ elements and is generated by the rotation $C_{nz}$ about the $z$ axis by an angle $2\pi/n$. 
The dihedral group $\mathbf{D}_n$ has $2n$ elements and is generated by $C_{nz}$ and also by a reflection $\sigma$ in a vertical plane. One can associate a ``parent'' 3D point group $\mathbf{C}_n$ or $\mathbf{C}_{nv}$ with each rosette group,
keeping in mind that, when acting on two-component vectors $\bk=(k_x,k_y)$ any reflection in a vertical plane is equivalent to a rotation by $\pi$ about a horizontal axis, hence the name ``dihedral''. For example, the reflection $\sigma_x$ 
in the $yz$ plane has the same effect as a $\pi$-rotation about the $y$ axis: $\sigma_x(k_x,k_y)=(-k_x,k_y)=C_{2y}(k_x,k_y)$. 

The states $|\bk,n\rangle$ and $K|\bk,n\rangle$ can be combined into a bispinor
\begin{equation}
\label{bispinor}
  |\Psi_{\bk,n}\rangle=\left(\begin{array}{c}
                           |\bk,n\rangle \\ K|\bk,n\rangle
                           \end{array}\right).
\end{equation}
To avoid double counting, the wavevectors labelling $|\Psi_{\bk,n}\rangle$ are restricted to a half of the Brillouin zone (HBZ), which should be chosen in such a way that it does not contain time-reversed momenta $\bk$ and $-\bk$ at the same time.
Since the electron bands are nondegenerate, action by a point group element $g$ on $|\bk,n\rangle$ produces a state which belongs to the wave vector $g\bk$ and differs from $|g\bk,n\rangle$ only by a phase factor: 
$g|\bk,n\rangle=e^{i\phi_{\bk,n}(g)}|g\bk,n\rangle$.
Using the commutation of $g$ and $K$ we have $gK|\bk,n\rangle=Kg|\bk,n\rangle=e^{-i\phi_{\bk,n}(g)}K|g\bk,n\rangle$, which leads to the following transformation properties: 
\begin{equation}
\label{trans-Psis}
    g|\Psi_{\bk,n}\rangle=\left(\begin{array}{cc}
                                e^{i\phi_{\bk,n}(g)} & 0 \\
                                0 & e^{-i\phi_{\bk,n}(g)}
                                \end{array}\right)|\Psi_{g\bk,n}\rangle,\quad 
    K|\Psi_{\bk,n}\rangle=\left(\begin{array}{cc}
                                0 & 1 \\
                                -1 & 0
                                \end{array}\right)|\Psi_{\bk,n}\rangle.  
\end{equation}
Note that, since the Bloch states $|\bk,n\rangle$ are spin-$1/2$ spinors, any rotation by $2\pi$ changes their sign. For instance, for the fourfold rotation $C_{4z}$ we have $C_{4z}^4|\bk,n\rangle=-|\bk,n\rangle$. 
This double-valuedness can be dealt with in the standard fashion, \cite{LL-3} 
by adding a fictitious new symmetry element $\bar E$ to the point group, which commutes with all other elements and satisfies the conditions $C_{4z}^4=\bar E$, $K^2=\bar E$, and $\bar E^2=E$, where $E$ is the identity element. 
Its action on the bispinors (\ref{bispinor}) is given by $\bar E|\Psi_{\bk,n}\rangle=-|\Psi_{\bk,n}\rangle$.

The transformation rules for the electron creation and annihilation operators are obtained from Eq. (\ref{trans-Psis}), if one views the Bloch states as vectors in the Fock space, \textit{i.e.}, 
$|\bk,n\rangle=\hat c^\dagger_{\bk,n}|0\rangle$, where $|0\rangle$ is the vacuum state, which is assumed to be invariant under all symmetry operations. Using Eq. (\ref{t_n}) we arrive at the following expression 
for the creation operator in the time-reversed state:
\begin{equation}
\label{c-tilde c}
  \hat{\tilde c}^\dagger_{\bk,n}\equiv K\hat c^\dagger_{\bk,n}K^{-1}=t_n(\bk)\hat c^\dagger_{-\bk,n},
\end{equation}
where $K^{-1}=K^\dagger$. Then,
\begin{eqnarray}
\label{trans-g-c}
  && g\hat c^\dagger_{\bk,n}g^{-1}=e^{i\phi_{\bk,n}(g)}\hat c^\dagger_{g\bk,n},\quad g\hat{\tilde c}^\dagger_{\bk,n}g^{-1}=e^{-i\phi_{\bk,n}(g)}\hat{\tilde c}^\dagger_{g\bk,n},\\
\label{trans-K-c}
  && K(f\hat c^\dagger_{\bk,n})K^{-1}=f^*\hat{\tilde c}^\dagger_{\bk,n},\quad K(f\hat{\tilde c}^\dagger_{\bk,n})K^{-1}=-f^*\hat c^\dagger_{\bk,n},
\end{eqnarray}
and also $\bar E\hat c^\dagger_{\bk,n}\bar E^{-1}=K\hat{\tilde c}^\dagger_{\bk,n}K^{-1}=-\hat c^\dagger_{\bk,n}$. We included a $c$-number factor $f$ in Eq. (\ref{trans-K-c}) to emphasize the antiunitarity of the TR operator.
The transformation rules for the annihilation operators are given by Hermitian conjugates of Eqs. (\ref{c-tilde c}), (\ref{trans-g-c}), and (\ref{trans-K-c}).

\subsection{Two-band model}
\label{sec: Rashba model}

To illustrate the points made above we start with the general Hamiltonian of noninteracting electrons in a TR invariant noncentrosymmetric crystal lattice with the SO coupling:\cite{Sam09}
\begin{equation}
\label{H pseudospin}
    \hat H_0=\sum_{\bk,\mu\nu}\sum_{\alpha,\beta=\uparrow,\downarrow}[\epsilon_\mu(\bk)\delta_{\mu\nu}\delta_{\alpha\beta}+
    iA_{\mu\nu}(\bk)\delta_{\alpha\beta}+\bm{B}_{\mu\nu}(\bk)\bm{\sigma}_{\alpha\beta}]\hat a^\dagger_{\bk\mu\alpha}\hat a_{\bk\nu\beta},
\end{equation}
where $\hat a^\dagger$ and $\hat a$ are the electron creation and annihilation operators in the pseudospin Bloch states, $\hat{\bm{\sigma}}$ are the Pauli matrices, and the wavevector summation is performed over the BZ. 
The first term, with $\epsilon_\mu(\bk)=\epsilon_\mu(-\bk)$, describes the twofold degenerate bands obtained from the inversion-symmetric lattice potential, including the intra-atomic SO coupling, which are labelled by the indices $\mu$ and $\nu$. 
All effects of the inversion-antisymmetric lattice potential and the corresponding SO coupling are contained in the last two terms. 

It follows from the requirements of lattice periodicity, Hermiticity, and TR invariance that $A_{\mu\nu}$ and $\bm{B}_{\mu\nu}$ are real, odd in $\bk$, periodic in reciprocal space, and satisfy $A_{\mu\nu}(\bk)=-A_{\nu\mu}(\bk)$ and 
$\bm{B}_{\mu\nu}(\bk)=\bm{B}_{\nu\mu}(\bk)$. The point group imposes additional constraints, see Ref. \onlinecite{Sam09} for details.
The pseudospin degeneracy of the bands is lifted if the $\bm{B}$ term is nonzero. However, the bands always remain twofold degenerate at the TR invariant points $\bk=\bK$, where $\bm{B}_{\mu\nu}(\bK)=\bm{0}$,
because $\bm{B}_{\mu\nu}(\bK)=-\bm{B}_{\mu\nu}(-\bK)=-\bm{B}_{\mu\nu}(\bK+\bm{G})=-\bm{B}_{\mu\nu}(\bK)$. For the same reason, $A_{\mu\nu}(\bK)=0$. Therefore, the diagonalization of Eq. (\ref{H pseudospin}) 
produces electronic bands that come in pairs connected at the TR invariant points. 

One can considerably simplify the problem and yet capture its essential physics by keeping just one such pair of bands, corresponding to $\mu=0$ (see Appendix \ref{sec: 4-band model} for a model with two pairs of bands).
Observing that $A_{00}(\bk)=0$ and introducing the notation $\bm{B}_{00}(\bk)=\bgam(\bk)$, 
we arrive at the following Hamiltonian:
\begin{equation}
\label{H-Rashba}
    \hat H_0=\sum\limits_{\bk,\alpha\beta}[\epsilon_0(\bk)\delta_{\alpha\beta}+\bgam(\bk)\bm{\sigma}_{\alpha\beta}]\hat a^\dagger_{\bk\alpha}\hat a_{\bk\beta}.
\end{equation}
The antisymmetric SO coupling is described by the 3D pseudovector $\bgam(\bk)$, which is real, odd in $\bk$ due to TR symmetry,
and invariant under the point group operations, \textit{i.e.,} $g\bgam(g^{-1}\bk)=\bgam(\bk)$ for any element $g$ of the 2D point group $G$. Recall that proper rotations act in the same way on the components of
the polar vector $\bk$ and the pseudovector $\bgam$, for instance, $C_{nz}k_\pm=e^{\pm 2\pi i/n}k_\pm$, where $k_\pm=k_x\pm ik_y$. In contrast, reflections act differently: $\sigma_x(k_x,k_y)=(-k_x,k_y)$, but 
$\sigma_x(\gamma_x,\gamma_y,\gamma_z)=(\gamma_x,-\gamma_y,-\gamma_z)$. Barring accidental degeneracies, all three components of $\bgam$ vanish simultaneously only at the TR invariant points,
where  $\bgam(\bK)=-\bgam(-\bK)=-\bgam(\bK+\bm{G})=-\bgam(\bK)=\bm{0}$.
The Hamiltonian of the form (\ref{H-Rashba}) is sometimes called the generalized Rashba model. In the original Rashba model, see Ref. \onlinecite{Rashba-model} and the references therein, the antisymmetric SO coupling of a particular form, 
$\bgam(\bk)=a(k_y\hat x-k_x\hat y)$, was used.

Diagonalization of Eq. (\ref{H-Rashba}) yields two bands of the form $\xi_\lambda(\bk)=\epsilon(\bk)+\lambda|\bgam(\bk)|$, where the band index $\lambda=\pm$ is called the helicity.
We will use the same term broadly to label the nondegenerate Bloch bands in the general case, \textit{i.e.}, beyond the generalized Rashba model.  
Using the spherical angle parametrization of the SO coupling, $\hat{\bgam}=\bgam/|\bgam|=(\sin\alpha\cos\beta,\sin\alpha\sin\beta,\cos\alpha)$, we obtain the following eigenstates:
\begin{equation}
\label{Bloch-Rashba}
    |\bk,+\rangle=\left(\begin{array}{c}
            \cos(\alpha/2)\\ e^{i\beta}\sin(\alpha/2)
                 \end{array}\right),\quad
    |\bk,-\rangle=\left(\begin{array}{c}
            \sin(\alpha/2)\\ -e^{i\beta}\cos(\alpha/2)
                 \end{array}\right).
\end{equation}
Physically, the helicity corresponds to the pseudospin projection on the direction of the SO coupling, because $\langle\bk,\lambda|\hat{\bm{\sigma}}|\bk,\lambda\rangle=\lambda\hat{\bgam}(\bk)$.
It follows from Eq. (\ref{t_n}) that in the generalized Rashba model the phase factor connecting the time-reversed helicity states is given by 
\begin{equation}
\label{t-factor-Rashba}
  t_\lambda(\bk)=\lambda e^{-i\beta(\bk)}.
\end{equation}
The expressions (\ref{Bloch-Rashba}) and (\ref{t-factor-Rashba}) are smooth and single valued everywhere in the BZ, except the TR invariant points, where the bands are degenerate and the angles $\alpha$ and $\beta$ are not defined. 

Representative expressions for $\bgam(\bk)$ near $\bk=\bm{0}$ for all 2D point groups are given in Table \ref{table gammas} (for the noncentrosymmetric 3D point groups, see Ref. \onlinecite{Sam09}). The reflection generator of the dihedral groups 
is assumed to be $\sigma_x$.
Note that for $G=\mathbf{C}_2$, $\mathbf{C}_4$, $\mathbf{C}_6$, $\mathbf{D}_2$, $\mathbf{D}_4$, and $\mathbf{D}_6$ the SO coupling is ``planar'' in the sense that $\bgam(\bk)$ lies in the $xy$ plane. This symmetry-imposed constraint is due
to the presence of the rotation $C_{2z}$ in the point group. Indeed, we have
$$
  \bgam(\bk)=C_{2z}\bgam(C^{-1}_{2z}\bk)=\left(\begin{array}{c}
            -\gamma_x(-\bk)\\ -\gamma_y(-\bk)\\ \gamma_z(-\bk)
                 \end{array}\right)=
         \left(\begin{array}{c}
            \gamma_x(\bk)\\ \gamma_y(\bk)\\ -\gamma_z(\bk)
                 \end{array}\right), 
$$
therefore $\gamma_z(\bk)=0$ at all $\bk$. Note that, although the antisymmetric SO coupling is not linear in $\bk$ in general, its planar component $\bgam_\parallel=(\gamma_x,\gamma_y,0)$ is,
which has important implications for the Bloch wave function topology, as we shall see in Sec. \ref{sec: Bloch Berry}.

To extend the expressions from Table \ref{table gammas} to the whole BZ and take into account the momentum space periodicity, $\bgam(\bk+\bm{G})=\bgam(\bk)$, the SO coupling can be represented as the lattice Fourier series:
\begin{equation}
\label{lattice-gamma}
  \bgam(\bk)=\sum_n\bgam_n\sin(\bk\bm{R}_n),
\end{equation}
where $\bgam_n$ are real pseudovectors and the summation goes over the sites $\bm{R}_n$ of a 2D Bravais lattice which cannot be transformed one into another by inversion. 
As examples we consider two simple 2D lattices, square and hexagonal, with the point groups $\mathbf{D}_4$ and $\mathbf{D}_6$, respectively. 
The former case describes, for instance, the crystalline symmetry of the 2D electron gas at the interface between two nonsuperconducting oxides.\cite{interface-SC}
In both cases, the lattice constant is equal to $d$ and we limit the summation in Eq. (\ref{lattice-gamma}) to the nearest-neighbor sites. 

For the square lattice, we obtain: 
\begin{equation}
\label{gamma-square}
  \bgam(\bk)=\gamma_0(\hat x\sin k_yd-\hat y\sin k_xd).
\end{equation}
The SO coupling vanishes at four TR invariant points given by 
\begin{equation}
\label{TRI-points-square}
  \{\bK_i\}=\left\{\bm{0},\ \frac{\bm{G}_1}{2},\ \frac{\bm{G}_2}{2},\ \frac{\bm{G}_1+\bm{G}_2}{2}\right\},
\end{equation}
where $\bm{G}_1=(2\pi/d)\hat x$ and $\bm{G}_2=(2\pi/d)\hat y$ are the primitive vectors of the reciprocal square lattice. For the hexagonal lattice, we obtain:
\begin{equation}
\label{gamma-hexagonal}
  \bgam(\bk)=\sqrt{3}\gamma_0\cos\left(\frac{k_xd}{2}\right)\sin\left(\frac{\sqrt{3}}{2}k_yd\right)\hat x
	      -\gamma_0\left[\sin(k_xd)+\sin\left(\frac{k_xd}{2}\right)\cos\left(\frac{\sqrt{3}}{2}k_yd\right)\right]\hat y.
\end{equation}
This expression has six zeros in the BZ, which are located at the following TR invariant points:
\begin{equation}
\label{TRI-points-hexagonal}
  \{\bK_i\}=\left\{\bm{0},\ \frac{\bm{G}_1}{2},\ \frac{\bm{G}_2}{2},\ \frac{\bm{G}_1-\bm{G}_2}{2},\ \frac{2\bm{G}_1-\bm{G}_2}{3},\ \frac{\bm{G}_1+\bm{G}_2}{3}\right\}, 
\end{equation}
where 
$$
  \bm{G}_1=\frac{2\pi}{d}\left(\hat x+\frac{1}{\sqrt{3}}\hat y\right),\quad \bm{G}_2=\frac{2\pi}{d}\frac{2}{\sqrt{3}}\hat y
$$  
are the primitive vectors of the reciprocal hexagonal lattice. It is easy to see that the expressions from Table \ref{table gammas} are recovered at small momenta in the vicinity of the $\Gamma$ point.

\begin{table}
\caption{Lowest-order polynomial expressions for the antisymmetric SO coupling near the center of the BZ ($a_i$ and $a$ are real
constants, $b$ is a complex constant, and $k_\pm=k_x\pm ik_y$). The ``parent'' 3D point groups are listed in the right column.}
\begin{tabular}{|c|c|c|}
    \hline
    \hspace*{2mm} $ G_{2D}$\hspace*{2mm}    & $\bgam(\bk)$ & $ G_{3D}$  \\ \hline
    $\mathbf{C}_1$      & $(a_1k_x+a_2k_y)\hat x+(a_3k_x+a_4k_y)\hat y+(a_5k_x+a_6k_y)\hat z$ & $\mathbf{C}_1$ \\ \hline
    $\mathbf{C}_2$      & $(a_1k_x+a_2k_y)\hat x+(a_3k_x+a_4k_y)\hat y$ & $\mathbf{C}_2$ \\ \hline
    $\mathbf{C}_3$    & $(a_1k_x+a_2k_y)\hat x+(-a_2k_x+a_1k_y)\hat y+(bk_+^3+b^*k_-^3)\hat z$ & $\mathbf{C}_3$ \\ \hline
    $\mathbf{C}_4$      & $(a_1k_x+a_2k_y)\hat x+(-a_2k_x+a_1k_y)\hat y$ & $\mathbf{C}_4$ \\ \hline
    $\mathbf{C}_6$     & $(a_1k_x+a_2k_y)\hat x+(-a_2k_x+a_1k_y)\hat y$ &  $\mathbf{C}_6$ \\ \hline
    $\mathbf{D}_1$      & $a_1k_y\hat x+a_2k_x\hat y+a_3k_x\hat z$ & \ $\mathbf{C}_{1v}\equiv\mathbf{C}_s$\hspace*{1mm} \\ \hline
    $\mathbf{D}_2$      & $a_1k_y\hat x+a_2k_x\hat y$ &  $\mathbf{C}_{2v}$ \\ \hline
    $\mathbf{D}_3$     & $a_1(k_y\hat x-k_x\hat y)+a_2(k_+^3+k_-^3)\hat z$ & $\mathbf{C}_{3v}$ \\ \hline
    $\mathbf{D}_4$       & $a(k_y\hat x-k_x\hat y)$ & $\mathbf{C}_{4v}$ \\ \hline
    $\mathbf{D}_6$     & $a(k_y\hat x-k_x\hat y)$ & $\mathbf{C}_{6v}$ \\ \hline
\end{tabular}
\label{table gammas}
\end{table}

\section{Superconductivity in nondegenerate bands}
\label{sec: SC}

We assume a BCS-like mechanism of superconductivity, in which the pairing interaction is only effective below a certain energy cutoff $\varepsilon_c$ near the 2D Fermi surface. The latter is represented by closed lines 
on the BZ torus. Below we use the notation $FS_n$ for the Fermi surface in the $n$th band, which is a set of points satisfying the equation $\xi_n(\bk)=0$.
In real noncentrosymmetric superconductors, one has the following hierarchy of the energy scales: $T_c\ll\varepsilon_c\ll E_{\mathrm{SO}},\epsilon_F$, where $E_{\mathrm{SO}}$ characterizes the SO band splitting near the Fermi energy. 
This means that the bands are sufficiently well separated to suppress the pairing of electrons with different helicities.

We construct the Hamiltonian using the basis of the exact band states $|\bk,n\rangle$, which include all effects of the lattice potential and the SO coupling. We have $\hat H=\hat H_0+\hat H_{int}$, where
\begin{equation}
\label{H_0}
     \hat H_0=\sum_{\bk,n}\xi_n(\bk)\hat c^\dagger_{\bk,n}\hat c_{\bk,n},
\end{equation}
describes noninteracting quasiparticles in the Bloch band and 
\begin{equation}
\label{H int}
    \hat H_{int}=\frac{1}{2{\cal V}}\sum\limits_{\bk\bk'\bq}\sum_{nn'}V_{nn'}(\bk,\bk')\hat c^\dagger_{\bk+\bq,n}\hat{\tilde c}^\dagger_{\bk,n}\hat{\tilde c}_{\bk',n'}\hat c_{\bk'+\bq,n'},
\end{equation}
is the pairing interaction between quasiparticles in the time-reversed states $|\bk,n\rangle$ and $K|\bk,n\rangle$ of the same helicity, see Eq. (\ref{c-tilde c}), and ${\cal V}$ is the system volume. It is legitimate
to neglect the dependence of the interaction strength on the center-of-mass momentum of the pairs, since $|\bq|$ is small compared to the Fermi momenta.
From Eqs. (\ref{trans-g-c}) and (\ref{trans-K-c}), and also the hermiticity of $H_{int}$, we obtain that
$\xi_n(\bk)=\xi_n(g^{-1}\bk)=\xi_n(-\bk)$ and
\begin{equation}
\label{tilde-V-symm}
  V_{nn'}(\bk,\bk')=V_{nn'}(g^{-1}\bk,g^{-1}\bk')=V^*_{nn'}(\bk,\bk')=V^*_{n'n}(\bk',\bk),
\end{equation}
\textit{i.e.}, the interaction functions are invariant under the point group and TR operations and independent of the phase choice for the Bloch states. 
The strength of intraband pairing is described by the diagonal elements $V_{nn}$, while that of the pair transfer between the bands -- by $V_{nn'}$ with $n\neq n'$.
The pairing interaction is nonzero only near the Fermi surfaces and the band summations in Eqs. (\ref{H_0}) and (\ref{H int}) is limited to 
$n=1,...,M$, where $M$ is the number of bands crossing the Fermi level and participating in superconductivity. 

One can decouple the pairing interaction (\ref{H int}) in the mean-field approximation and obtain: 
\begin{equation}
\label{H_mf}
     \hat H_{int}=\sum_{\bk\in\mathrm{HBZ}}\sum_{n=1}^M\left[\Delta_n(\bk)\hat c^\dagger_{\bk,n}\hat{\tilde c}^\dagger_{\bk,n}+\Delta^*_n(\bk)\hat{\tilde c}_{\bk,n}\hat c_{\bk,n}\right],
\end{equation}
where $\Delta_n(\bk)$ is the gap function in the $n$th band. 
The interband gap functions are equal to zero, due to the large SO band splitting. It follows from Eq. (\ref{c-tilde c}) and the anticommutation of the fermion creation and annihilation operators that
\begin{equation}
\label{Delta-even}
  \Delta_n(\bk)=\Delta_n(-\bk).
\end{equation}
We omitted a $c$-number term in Eq. (\ref{H_mf}) and restricted the wavevector summation to the half of the BZ, see the discussion after Eq. (\ref{bispinor}). 
Similarly, the noninteracting Hamiltonian can be written as
\begin{equation}
\label{H_0-ctc}
     \hat H_0=\sum_{\bk\in\mathrm{HBZ}}\sum_{n=1}^M\left[\xi_n(\bk)\hat c^\dagger_{\bk,n}\hat c_{\bk,n}+\xi_n(\bk)\hat{\tilde c}^\dagger_{\bk,n}\hat{\tilde c}_{\bk,n}\right].
\end{equation}
Next, we introduce two-component Nambu creation and annihilation operators associated with each $\bk\in\mathrm{HBZ}$:
\begin{equation}
\label{Nambu-C-def}
  \hat C_{\bk,n}=\left(\begin{array}{c}
                        \hat c_{\bk,n} \\ 
                        \hat{\tilde c}^\dagger_{\bk,n}
                       \end{array}\right), \quad \hat C^\dagger_{\bk,n}=\left(\hat c^\dagger_{\bk,n},\hat{\tilde c}_{\bk,n}\right),
\end{equation} 
and combine Eqs. (\ref{H_mf}) and (\ref{H_0-ctc}) into the following expression:
\begin{equation}
\label{H-BdG_1}
  \hat H=\sum_{\bk\in\mathrm{HBZ}}\sum_{n=1}^M\hat C^\dagger_{\bk,n}\hat h_n(\bk)\hat C_{\bk,n},
\end{equation}
where
\begin{equation}
\label{h-def}
  \hat h_n(\bk)=\left(\begin{array}{cc}
           \xi_n(\bk) & \Delta_n(\bk) \\
           \Delta_n^*(\bk) & -\xi_n(\bk)
          \end{array}\right)=\hat h_n(-\bk)
\end{equation}
is called the Nambu Hamiltonian. Due to the two-component nature of the Nambu wave functions, which mix the electron- and hole-like quasiparticle states, the total number of degrees of freedom is unchanged compared to the noninteracting Hamiltonian, 
despite the halving of the number of independent wavevectors.  

It follows from Eq. (\ref{tilde-V-symm}) that the gap functions transform like complex scalars under the point group and TR operations, namely,
\begin{equation}
\label{trans-Delta}
  g:\ \Delta_n(\bk) \to \Delta_n(g^{-1}\bk),\qquad K:\ \Delta_n(\bk) \to \Delta^*_n(\bk)
\end{equation}
(there is no need for the double group anymore, since $\bar E$ is equivalent to the identity element $E$ when acting on the gap functions). These transformation properties are consequences of the Cooper pairs being built from the time-reversed
states $|\bk,n\rangle$ and $K|\bk,n\rangle$, as opposed to $|\bk,n\rangle$ and $|-\bk,n\rangle$. In the latter case, the mean-field gap functions in Eq. (\ref{H_mf}) are replaced by $\tilde\Delta_n(\bk)=t_n(\bk)\Delta_n(\bk)$, 
which do not have any simple symmetry properties.\cite{SCSam-04} Furthermore, we will see in Sec. \ref{sec: BdG eqs} that it is $\Delta_n$, not $\tilde\Delta_n$, which enters the proper BdG Hamiltonian. 
Note that advantages of pairing up the time-reversed states have long been recognized in centrosymmetric superconductors, see Ref. \onlinecite{And84}. 

Due to the properties (\ref{trans-Delta}), the gap function in each band can be represented as a linear combination of the basis functions of the irreducible representations (IREPs) of the point group $G$. 
In the pairing channel corresponding to the IREP $\Gamma$ of dimensionality $d_\Gamma$, we have
\begin{equation}
\label{Delta expansion}
    \Delta_n(\bk)=\sum_{a=1}^{d_{\Gamma}}\eta_{n,a}\phi_a(\bk).
\end{equation}
Here $Md_\Gamma$ complex coefficients $\eta_{n,a}$ play the role of the order parameter components and $\phi_a(\bk)$ are the basis functions. In a BCS-like model, the gap function in the $n$th band is nonzero only if $|\xi_n(\bk)|\leq\varepsilon_c$. 
For brevity, the momentum cutoff factors are not explicitly shown in Eq. (\ref{Delta expansion}).

It follows from Eq. (\ref{Delta-even}) that the basis functions that appear in the expansion of $\Delta_n(\bk)$ have to be even in $\bk$. 
In general, they can have different momentum dependence in different bands, which is neglected in Eq. (\ref{Delta expansion}) and everywhere below. 
In Tables \ref{table IREPs cyclic} and \ref{table IREPs dihedral} we list the IREPs for all 2D point groups, together with the lowest-order polynomial expressions for their even basis functions near the center of the BZ.
The basis functions of all $A$ and $B$ IREPs can be chosen to be real. The cyclic groups $\mathbf{C}_3$, $\mathbf{C}_4$, and $\mathbf{C}_6$ have pairs of complex conjugate one-dimensional (1D) representations, labelled by $E^{(\pm)}$. In the 
absence of TR symmetry breaking in the normal state, any such pair of IREPs should be treated as one 2D ``physically irreducible'' representation,\cite{LL-3} with the corresponding order parameters having two components. 
Note that certain pairing channels cannot be realized, because the corresponding IREPs do not have even basis functions, which can be understood as follows. If a 2D point group contains the twofold rotation about 
the $z$-axis, $C_{2z}(k_x,k_y)=(-k_x,-k_y)$, then its character in the IREPs labelled with a dash in Tables \ref{table IREPs cyclic} and \ref{table IREPs dihedral} 
is equal to $-1$ in the 1D IREPs and $-2$ in the 2D IREPs,\cite{LL-3} therefore the basis functions have to odd in $\bk$. 

The order parameter cannot have more than two components in each band, because all IREPs are either 1D or 2D. The stable uniform superconducting states are found by minimizing the Ginzburg-Landau free energy $F[\eta_1,...,\eta_M]$ in the 1D case,
while in the 2D case the free energy has the form $F[\bm{\eta}_1,...,\bm{\eta}_M]$, where $\bm{\eta}_n=(\eta_{n,1},\eta_{n,2})$. Since the values of the order parameter components are not important for our discussion below, 
we will not attempt the free energy minimization here. One can expect that it can be done explicitly only in the simplest models. For example, in the
case of an 1D IREP with $M=2$, the free energy is formally the same as in the usual two-band BCS model, which has recently attracted a lot of attention due to its applications to MgB$_2$ (Ref. \onlinecite{MgB2-review}), 
iron-based high-temperature superconductors,\cite{Fe-based} and other materials. In addition to the TR invariant states, in which the phase difference between the order parameters $\eta_1$ and $\eta_2$ is either $0$ or $\pi$,   
there are also stable states that break TR invariance.\cite{Sam15}

\begin{table}
\caption{Even basis functions of the IREPs of the cyclic 2D point groups ($a$ and $b$ are real constants, and $k_\pm=k_x\pm ik_y$).}
\begin{tabular}{|c|c|c|c|}
    \hline
    \hspace*{3mm} $G$\hspace*{3mm}   & \hspace*{4mm}$\Gamma$\hspace*{4mm} & \hspace*{1mm} $d_\Gamma$\hspace*{1mm} & $\phi_\Gamma(\bk)=\phi_\Gamma(-\bk)$ \\ \hline
    $\mathbf{C}_1$   & $A$ & $1$ & $1$ \\ \hline
    $\mathbf{C}_2$   & $A$ & $1$ & $1$ \\ 
		  	& $B$ & $1$ & $-$ \\ \hline
    $\mathbf{C}_3$    & $A$ & $1$ & $1$ \\ 
                    & $E^{(\pm)}$ & $1$ & $k_\pm^2$ \\ \hline
    $\mathbf{C}_4$    & $A$ & $1$ & $1$ \\ 
		      & $B$ & $1$ & $a(k_x^2-k_y^2)+bk_xk_y$ \\ 
		      & $E^{(\pm)}$ & $1$ & $-$ \\ \hline
    $\mathbf{C}_6$    & $A$ & $1$ & $1$ \\
		      & $B$ & $1$ & $-$ \\
		      & $E_1^{(\pm)}$ & $1$ & $k_\pm^2$ \\
		      & $E_2^{(\pm)}$ & $1$ & $-$ \\ \hline
\end{tabular}
\label{table IREPs cyclic}
\end{table}

\begin{table}
\caption{Even basis functions of the IREPs of the dihedral 2D point groups ($k_\pm=k_x\pm ik_y$).}
\begin{tabular}{|c|c|c|c|}
    \hline
    \hspace*{3mm} $G$\hspace*{3mm}   & \hspace*{4mm}$\Gamma$\hspace*{4mm} & \hspace*{1mm} $d_\Gamma$\hspace*{1mm} & $\phi_\Gamma(\bk)=\phi_\Gamma(-\bk)$ \\ \hline
    $\mathbf{D}_1$    & $A_1$ & $1$ & $1$ \\
                      & $A_2$ & $1$ & $k_xk_y$ \\ \hline
    $\mathbf{D}_2$    & $A$ & $1$ & $1$ \\ 
                      & $B_1$ & $1$ & $k_xk_y$ \\ 
                      & $B_2$ & $1$ & $-$ \\ 
                      & $B_3$ & $1$ & $-$ \\ \hline
    $\mathbf{D}_3$   & $A_1$ & $1$ & $1$ \\ 
                     & $A_2$ & $1$ & $i(k_+^6-k_-^6)$ \\ 
                     & $E$ & $2$ & $k_+(k_+^3-k_-^3),\ k_-(k_+^3-k_-^3)$ \\ \hline
    $\mathbf{D}_4$    & $A_1$ & $1$ & $1$ \\ 
                      & $A_2$ & $1$ & $k_xk_y(k_x^2-k_y^2)$ \\ 
                      & $B_1$ & $1$ & $k_x^2-k_y^2$ \\ 
                      & $B_2$ & $1$ & $k_xk_y$ \\ 
                      & $E$ & $2$ & $-$ \\ \hline
    $\mathbf{D}_6$    & $A_1$ & $1$ & $1$ \\ 
                      & $A_2$ & $1$ & $i(k_+^6-k_-^6)$ \\ 
                      & $B_1$ & $1$ & $-$ \\ 
                      & $B_2$ & $1$ & $-$ \\ 
                      & $E_1$ & $2$ & $-$ \\ 
                      & $E_2$ & $2$ & $k_+^2,\ k_-^2$ \\ \hline
\end{tabular}
\label{table IREPs dihedral}
\end{table}

\subsection{Bogoliubov-de Gennes Hamiltonian}
\label{sec: BdG eqs}

The energies and wave functions of fermionic quasiparticles in the superconducting state are given by the eigenvalues and eigenfunctions of a certain first-quantization operator, known as the BdG Hamiltonian. 
It can be constructed by considering an arbitrary basis of single-particle 
Bloch states $|\bk,i\rangle$ at each $\bk$, labelled by some quantum numbers $i$. For instance, in the generalized Rashba model, see Eq. (\ref{H-Rashba}), one can use the pseudospin basis with $i=\alpha=\uparrow,\downarrow$. 
The second quantization operators are transformed into the new basis according to
$$
  \hat c_{\bk,n}=\sum_i\langle\bk,n|\bk,i\rangle\hat a_{\bk,i},\quad\hat c^\dagger_{\bk,n}=\sum_i\langle\bk,i|\bk,n\rangle\hat a_{\bk,i}^\dagger,
$$
while for the creation and annihilation operators in the time-reversed states we obtain: 
$$
  \hat{\tilde c}_{\bk,n}=\sum_i\langle\bk,i|\bk,n\rangle\hat{\tilde a}_{\bk,i},\quad\hat{\tilde c}^\dagger_{\bk,n}=\sum_i\langle\bk,n|\bk,i\rangle\hat{\tilde a}_{\bk,i}^\dagger,  
$$
where $\hat{\tilde a}_{\bk,i}=K\hat a_{\bk,i}K^{-1}$ and $\hat{\tilde a}_{\bk,i}^\dagger=K\hat a_{\bk,i}^\dagger K^{-1}$. Therefore, both components of the Nambu operators (\ref{Nambu-C-def}) transform in the same way 
and Eq. (\ref{H-BdG_1}) takes the following form:
\begin{equation}
\label{H-BdG_2} 
  \hat H=\sum_{\bk\in\mathrm{HBZ}}\sum_{ij}
  (\hat a^\dagger_{\bk,i},\hat{\tilde a}_{\bk,i})
  \langle\bk,i|{\cal H}_{BdG}(\bk)|\bk,j\rangle
  \left(\begin{array}{c}
         \hat a_{\bk,j} \\ \hat{\tilde a}^\dagger_{\bk,j}
        \end{array}\right).
\end{equation}
Here
\begin{equation}
\label{H-BdG}
  {\cal H}_{BdG}(\bk)=\sum_n |\bk,n\rangle \hat h_n(\bk)\langle\bk,n|=\sum_n\hat\Pi_n(\bk)\otimes\hat h_n(\bk),
\end{equation}
$\hat h_n$ is given by Eq. (\ref{h-def}), and 
$$
  \hat\Pi_n(\bk)=|\bk,n\rangle\langle\bk,n|
$$ 
is the projector onto the $n$th Bloch band. Thus, we have represented the mean-field BCS Hamiltonian in an arbitrary basis in terms of the matrix elements of an operator, called the BdG Hamiltonian, which acts in the 
$2M$-dimensional Hilbert space $L_B\times L_N$, where $L_B$ is the $M$-dimensional space of single-particle Bloch states and $L_N$ is the 2D Nambu space. 
Note that the BdG Hamiltonian with the correct transformation properties necessarily has $\Delta_n(\bk)$, not $t_n(\bk)\Delta_n(\bk)$, as its off-diagonal Nambu matrix elements, therefore the ambiguity of choosing 
the gap function in the BdG equations in the helicity basis, see, \textit{e.g.}, Ref. \onlinecite{SF09}, is avoided. 

At each $\bk$, the BdG Hamiltonian (\ref{H-BdG}) has $2M$ eigenstates 
\begin{equation}
\label{Phi-ns}
  \Phi_{n,s}(\bk)=|\bk,n\rangle\otimes|\bk,n;s\rangle, 
\end{equation}
where $s$ labels the electron-like ($s=+$) or hole-like ($s=-$) branches of the spectrum and $|\bk,n;s\rangle$ are eigenstates of the Nambu Hamiltonian $\hat h_n(\bk)$. The corresponding eigenvalues are given by $sE_n(\bk)$, where 
\begin{equation}
\label{E-n}
  E_n(\bk)=\sqrt{\xi_n^2(\bk)+|\Delta_n(\bk)|^2}
\end{equation}
is the energy of the Bogoliubov fermionic excitations in the $n$th band. As functions of $\bk$, the eigenvalues form $2M$ Bogoliubov bands in momentum space. 
The gap in the excitation energy is given by $|\Delta_n(\bk)|$, which can vanish at some high-symmetry points or lines in the BZ, see Tables \ref{table IREPs cyclic} and \ref{table IREPs dihedral}. A line of zeros crossing
a Fermi surface produces a point-like gap node. For instance, the superconducting states corresponding to the non-identity 1D IREPs of $\mathbf{D}_4$ all have gap nodes: at $k_x=\pm k_y$ for $\Gamma=B_1$, at $k_x=0$ and $k_y=0$ for $\Gamma=B_2$,
at $k_x=0$, $k_y=0$, and $k_x=\pm k_y$ for $\Gamma=A_2$. 

When expressed in terms of the excitation energy, the Nambu Hamiltonian takes the following form:
\begin{equation}
\label{h-nu}
  \hat h_n(\bk)=\bm{\nu}_n(\bk)\hat{\bm{\tau}}=E_n(\bk)\hat{\bm{\nu}}_n(\bk)\hat{\bm{\tau}},
\end{equation}
where $\hat{\bm{\tau}}$ are the Pauli matrices in the Nambu space,
$$
  \bm{\nu}_n(\bk)=\left(\begin{array}{c}
                         \re\Delta_n(\bk)\\
                         -\im\Delta_n(\bk)\\
                         \xi_n(\bk)
                        \end{array}\right),
$$
and $\hat{\bm{\nu}}_n=\bm{\nu}_n/|\bm{\nu}_n|$ is a unit vector. Due to the properties (\ref{xi-even}) and (\ref{Delta-even}), both 
$E_n$ and ${\bm{\nu}}_n$ are even functions of $\bk$. If the excitations are fully gapped, then $\hat{\bm{\nu}}_n$ is well defined everywhere in the BZ.

\section{Green's function topological invariants}
\label{sec: GFs}

As shown in the previous section, with each wavevector in the half of the BZ of a noncentrosymmetric superconductor one can associate the BdG Hamiltonian, which is given by a $2M\times 2M$ Hermitian matrix of a 
special form, see Eq. (\ref{H-BdG}). Therefore, the superconducting states can be classified into different universality classes, 
according to the topology of the mapping 
\begin{equation}
\label{k to H map}
  \bk\to{\cal H}_{BdG}(\bk).
\end{equation}
These universality classes are characterized by topological invariants, which are obtained by integrating certain differential forms constructed from
${\cal H}_{BdG}$ over closed manifolds in momentum space. Since we focus on 2D superconductors, the appropriate integration domains are either the 2D BZ itself or its
submanifolds. The topology of 3D noncentrosymmetric superconductors was studied in Ref. \onlinecite{Sam15}.

One possible way of enumerating the universality classes of the mapping (\ref{k to H map}) is based on the Maurer-Cartan topological invariants, see Ref. \onlinecite{Stone-book} and Appendix \ref{sec: MC-inv}. 
Since the Maurer-Cartan invariants are nonzero only in odd dimensions, we introduce, following Ref. \onlinecite{Volovik-book}, an auxiliary real variable $k_0$ (``frequency') and 
define the BdG Green's function as follows: ${\cal G}(\bk,k_0)=[ik_0-{\cal H}_{BdG}(\bk)]^{-1}$. Using Eqs. (\ref{H-BdG}) and (\ref{h-nu}), we obtain:
\begin{equation}
\label{GF-gen}
  {\cal G}(\bk,k_0)=\sum_n\hat\Pi_n(\bk)\otimes\hat g_n(\bk),
\end{equation}
where
$$
  \hat g_n(\bk,k_0)=-\frac{ik_0+\bm{\nu}_n(\bk)\hat{\bm{\tau}}}{k_0^2+E_n^2(\bk)}.
$$
The Maurer-Cartan invariants are constructed using the $2M\times 2M$ matrix-valued 1-form $\omega={\cal G}d{\cal G}^{-1}$. We assume a fully gapped spectrum, therefore ${\cal G}$ is nonsingular and the Maurer-Cartan form is well defined everywhere in the BZ. 
We focus on the following invariant:
\begin{equation}
\label{MC-I3}
  I_{2+1}=\int\Tr\omega^3,
\end{equation}
where ``$\Tr$'' stands for $2M\times 2M$ matrix trace and combined matrix and exterior multiplication is implied in $\omega^3$. Although the BdG Hamiltonian and the 1-form $\omega$ are defined only in the HBZ, one can show that 
$\Tr\omega^D(-\bk,k_0)=\Tr\omega^D(\bk,k_0)$ for all $D$, due to TR symmetry. Therefore, $\Tr\omega^3$ can be extended to the whole BZ and integrated over a closed $(2+1)$-dimensional manifold with coordinates $k_x,k_y,k_0$, 
which is topologically equivalent to a 3D torus (the frequency variable runs over the whole real axis, which is assumed to be closed into a circle).

From Eq. (\ref{GF-gen}) we obtain the following expression for the Maurer-Cartan form:
\begin{equation}
\label{omega-PQ}
  \omega=\sum_{mn}\hat\Pi_md\hat\Pi_n\otimes \hat P_{mn}+\sum_n\hat\Pi_n\otimes \hat Q_n,
\end{equation}
where 
\begin{equation}
\label{PQ}
  \hat P_{mn}=a_{mn}+i\bm{b}_{mn}\hat{\bm{\tau}},\quad \hat Q_n=c_n+i\bm{d}_n\hat{\bm{\tau}}
\end{equation}
are $2\times 2$ matrix-valued $0$- and $1$-forms, respectively, with 
\begin{eqnarray*}
  && a_{mn}=\frac{k_0^2+\bm{\nu}_m\bm{\nu}_n}{k_0^2+E_m^2},\quad \bm{b}_{mn}=\frac{\bm{\nu}_m\times\bm{\nu}_n-k_0(\bm{\nu}_m-\bm{\nu}_n)}{k_0^2+E_m^2}\\
  && c_n=\frac{k_0dk_0+\bm{\nu}_nd\bm{\nu}_n}{k_0^2+E_n^2},\quad \bm{d}_n=\frac{\bm{\nu}_n\times d\bm{\nu}_n+k_0d\bm{\nu}_n-dk_0\,\bm{\nu}_n}{k_0^2+E_n^2}. 
\end{eqnarray*}

To facilitate the calculation of the topological invariant (\ref{MC-I3}) we introduce the Bloch Berry connections in momentum space:\cite{Berry-phase-book}
\begin{equation}
\label{Berry-A-form}
  {\cal A}_{mn}(\bk)=i\langle\bk,m|d|\bk,n\rangle,
\end{equation}
which form a Hermitian $M\times M$ matrix-valued $1$-form. Then,
\begin{equation}
\label{Tr-omega3-K}
  \Tr\omega^3=\sum_nK^{(0)}_n+\sum_{mn}K^{(2)}_{mn}{\cal A}_{mn}{\cal A}_{nm}+\sum_{lmn}K^{(3)}_{lmn}{\cal A}_{lm}{\cal A}_{mn}{\cal A}_{nl},
\end{equation}
where
\begin{eqnarray*}
  K^{(0)}_n &=& 2[c_n^3-c_n(\bm{d}_n\bm{d}_n)+\bm{d}_n(\bm{d}_n\times\bm{d}_n],\\
  K^{(2)}_{mn} &=& -3(a_{mn}a_{nm}-\bm{b}_{mn}\bm{b}_{nm}-a_{mn}-a_{nm}+1)(c_m-c_m)\\
  && +3(a_{mn}\bm{b}_{nm}+a_{nm}\bm{b}_{mn}-\bm{b}_{mn}-\bm{b}_{nm})(\bm{d}_m-\bm{d}_n),\\
  K^{(3)}_{lmn} &=& 2i[a_{lm}a_{mn}a_{nl}-a_{lm}(\bm{b}_{mn}\bm{b}_{nl})-a_{mn}(\bm{b}_{lm}\bm{b}_{nl})-a_{nl}(\bm{b}_{lm}\bm{b}_{mn})+\bm{b}_{lm}(\bm{b}_{mn}\times\bm{b}_{nl})\\
  && -a_{lm}a_{mn}-a_{mn}a_{nl}-a_{nl}a_{lm}+(\bm{b}_{lm}\bm{b}_{mn})+(\bm{b}_{mn}\bm{b}_{nl})+(\bm{b}_{nl}\bm{b}_{lm})\\
  && +a_{lm}+a_{mn}+a_{nl}-1]
\end{eqnarray*}
are $3$-, $1$-, and $0$-forms, respectively. The derivation of these expressions is outlined in Appendix \ref{sec: long derivation}. 
The third term on the right-hand side of Eq. (\ref{Tr-omega3-K}) vanishes upon $(2+1)$-dimensional integration, because the Berry connections do not contain $k_0$. In the second term, one can write $K^{(2)}_{mn}=\rho dk_0+(...)$, where 
the ellipsis stands for the terms that do not contribute to the integral. A straightforward calculation gives $\rho=0$, therefore the second term also vanishes. The only remaining term can be simplified and becomes  
$$
  K^{(0)}_n=-\frac{6dk_0}{(k_0^2+E_n^2)^2}\bm{\nu}_n(d\bm{\nu}_n\times d\bm{\nu}_n).
$$
This expression is nonzero only near the $n$th Fermi surface, because $\bm{\nu}_n=\xi_n\hat z$ outside the BCS momentum shell. Integrating over the frequency variable $k_0$ in Eq. (\ref{MC-I3}), we obtain:
\begin{equation}
\label{I-3-nu}
  I_{2+1}=-3\pi\sum_n\int_{BZ}\hat{\bm{\nu}}_n(d\hat{\bm{\nu}}_n\times d\hat{\bm{\nu}}_n).
\end{equation}
Note that the Maurer-Cartan invariant depends only on the details of the Nambu Hamiltonian $\hat h_n(\bk)$, \textit{i.e.}, 
on the quasiparticle band energies and the superconducting gap functions, while all information about the topology of the normal-state Bloch bands, which is contained in the band projectors and the Berry connections, has been lost.

Further progress can be achieved for a fully gapped superconducting state of the form
\begin{equation}
\label{nodeless gap}
  \Delta_n(\bk)=|\Delta_n(\bk)|e^{i\varphi_n(\bk)},
\end{equation}
where the gap amplitudes do not vanish anywhere at the Fermi surface. Inserting this into Eq. (\ref{I-3-nu}) and integrating over $\xi_n$, we finally obtain:
$$ 
  I_{2+1}=-24\pi^2\sum_n N_n,
$$
where
\begin{equation}
\label{winding-number}
  N_n=\frac{1}{2\pi}\oint_{FS_n}d\varphi_n
\end{equation}
is the phase winding number of the gap function along the $n$th Fermi surface. The latter's orientation is fixed by $\hat z\times\bm{v}_{F,n}$, where $\bm{v}_{F,n}$ is the Fermi velocity, therefore the integration in
Eq. (\ref{winding-number}) is performed counterclockwise for electron-like Fermi surfaces and clockwise for hole-like Fermi surfaces. To make the topological invariant integer valued, we introduce
\begin{equation}
\label{N-3}
  \tilde N\equiv-\frac{1}{24\pi^2}I_{2+1}=\sum_n N_n,
\end{equation}
therefore the Maurer-Cartan invariant is nothing but the total phase winding number of the gap functions.

For the order parameter corresponding to a 1D IREP of the point group, the expression (\ref{N-3}) vanishes, but one can get a nonzero phase winding in TR breaking superconducting states with intrinsically complex gap functions. 
For example, choosing real basis functions for the 2D IREP $E_2$ of $\mathbf{D}_6$, we have from Eq. (\ref{Delta expansion}) and Table \ref{table IREPs dihedral}: $\Delta_n(\bk)=\eta_{n,1}(k_x^2-k_y^2)+2\eta_{n,2}k_xk_y$. 
If there is a stable superconducting state of the form $\bm{\eta}_n\propto(1,i)$, called the chiral $d$-wave state, then 
\begin{equation}
\label{chiral-d-wave}
  \Delta_n(\bk)=\Delta_{n,0}(\hat k_x^2-\hat k_y^2+2i\hat k_x\hat k_y),
\end{equation}
where $\hat\bk=\bk/|\bk|$. The phase winding number is equal to $2$ for any electron-like Fermi surface enclosing the $\Gamma$ point in the BZ, therefore $\tilde N=2M$ [for $\bm{\eta}_n\propto(1,-i)$ one has $N_n=-2$ and $\tilde N=-2M$].
Similar results are obtained for the chiral $d$-wave states corresponding to the ``physically irreducible'' 2D representations $E^{(\pm)}$ of $\mathbf{C}_3$, or $E^{(\pm)}_1$ of $\mathbf{C}_6$. 

Another example of a topologically nontrivial state is the $d+id$ state with $\Delta_n(\bk)=\Delta_{n,1}(\hat k_x^2-\hat k_y^2)+2i\Delta_{n,2}\hat k_x\hat k_y$ ($\Delta_{n,1}$ and $\Delta_{n,2}$ are assumed to be real), 
which is produced by mixing two 1D IREPs $B_1$ and $B_2$ of $\mathbf{D}_4$. The phase winding numbers are equal to $\pm 2$, depending on the relative sign of the order parameter components,\cite{Vol97}
and the topological invariant (\ref{N-3}) is given by $\tilde N=2\sum_n\sgn(\Delta_{n,1}\Delta_{n,2})$. Note that the chiral $p$-wave states with $\Delta_n(\bk)\propto k_x\pm ik_y$ are not possible, because the gap 
functions in the band representation are even in momentum.

\subsection{Boundary modes}
\label{sec: ABS}

If the bulk superconducting state is topologically nontrivial, then one can expect that gapless fermionic modes exist near the system's boundary.\cite{Volovik-book}
In this subsection we derive the spectrum of the boundary modes in the helicity band representation and establish its relation to the phase winding numbers, see Eq. (\ref{winding-number}). 
We consider a superconductor occupying the $x\geq 0$ half space, with a specularly reflecting surface. To make analytical progress, we neglect self-consistency and assume that the order parameter is uniform. 

For simplicity, we consider the case of $M=1$, so that there is just one Bloch band participating in superconductivity, which allows us to drop the band index. Due to the translational invariance along the surface, $k_y$ is a good quantum number.
The Bogoliubov quasiparticle spectrum at given $k_y$ can be found using the semiclassical (Andreev) approximation,\cite{And64} in which the wave function is sought in the form $e^{i\bk_F\br}\psi_{\bk_F}(x)$, where 
$\bk_F=(k_{F,x},k_y)$ is a Fermi wavevector and $k_{F,x}$ is a root of the equation
\begin{equation}
\label{FS-intersect-eq}
	\xi(k_{F,x},k_y)=0.
\end{equation}
The two-component Nambu spinor $\psi_{\bk_F}$ varies slowly on the scale of the Fermi wavelength and satisfies the Andreev equation:  
\begin{equation}
\label{And-eq-gen}
	\left(\begin{array}{cc}
		-iv_{F,x}\nabla_x & \Delta(\bk_F) \\
		\Delta^*(\bk_F) & iv_{F,x}\nabla_x
	\end{array}\right)\psi_{\bk_F}=E\psi_{\bk_F},
\end{equation}
where $v_{F,x}$ is the $x$-projection of the Fermi velocity at $\bk_F$. Thus, for a given momentum along the surface, each root of Eq. (\ref{FS-intersect-eq}) defines a semiclassical trajectory, and the Andreev wave function 
associated with this trajectory is found from Eq. (\ref{And-eq-gen}). Depending on the band structure, Eq. (\ref{FS-intersect-eq}) can have several solutions. 
The corresponding trajectories are classified as either incident ($v_{F,x}<0$) or reflected ($v_{F,x}>0$). For the trajectories parallel to the surface, we have $v_{F,x}=0$ and the semiclassical approximation is not applicable.  

We assume a fully gapped superconducting state of the form (\ref{nodeless gap}) with a constant gap magnitude $|\Delta(\bk)|=\Delta_0$.
Focusing on the quasiparticle states localized near the surface, which are called the Andreev bound states (ABS), we expect that $|E|\leq\Delta_0$. The ABS solution of Eq. (\ref{And-eq-gen}) has the form 
$\psi_{\bk_F}(x)=\phi(\bk_F)e^{-\Omega x/|v_{F,x}|}$, where
\begin{equation}
\label{Andreev amplitude}
	\phi(\bk_F)\equiv\psi_{\bk_F}(x=0)=C(\bk_F)\left(\begin{array}{c}
		\dfrac{\Delta(\bk_F)}{E-i\Omega\sgn v_{F,x}} \\ 1
	\end{array}\right)
\end{equation}
and $\Omega=\sqrt{\Delta_0^2-E^2}$ and $C$ is a coefficient. 

The semiclassical approximation breaks down near the surface due to the rapid variation of the lattice potential. Following Ref. \onlinecite{Shel-bc}, we describe the effects of the surface scattering by an effective boundary condition at $x=0$, 
which expresses the Andreev wave functions (\ref{Andreev amplitude}) for the reflected trajectories in terms of those for the incident trajectories. In the general case, if Eq. (\ref{FS-intersect-eq}) has an 
equal number (${\cal N}$) of incident and reflected solutions, corresponding to the Fermi wavevectors $\bk_{\mathrm{in},1},...,\bk_{\mathrm{in},{\cal N}}$ and $\bk_{\mathrm{out},1},...,\bk_{\mathrm{out},{\cal N}}$, respectively, 
the boundary condition can be written in the following form:  
\begin{equation}
\label{Shelankov-bc}
  \phi(\bk_{\mathrm{out},i})=\sum_{j=1}^{{\cal N}} S_{ij}\phi(\bk_{\mathrm{in},j}).
\end{equation}
Here $\hat S$ is an ${\cal N}\times{\cal N}$ unitary matrix, which is determined by the microscopic details of the surface scattering in the normal state.
Inserting the wave functions (\ref{Andreev amplitude}) into Eq. (\ref{Shelankov-bc}), we obtain a homogeneous system of $2{\cal N}$ linear equations for the coefficients $C(\bk_{\mathrm{in},1}),...,C(\bk_{\mathrm{in},{\cal N}})$
and $C(\bk_{\mathrm{out},1}),...,C(\bk_{\mathrm{out},{\cal N}})$. Equating its determinant to zero yields an equation for the ABS energy $E$. 

\begin{figure}
\includegraphics[width=8cm]{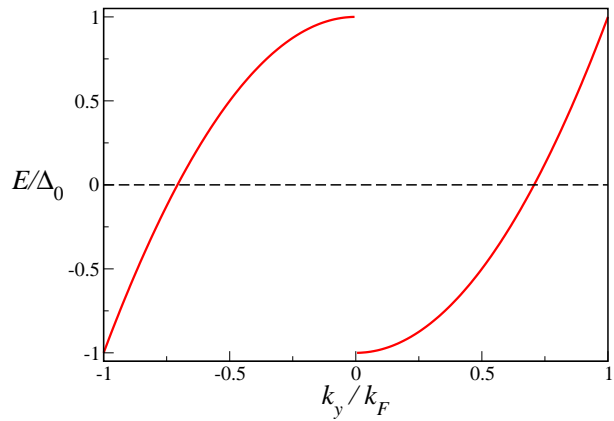}
\caption{(Color online) The ABS energy for the chiral $d$-wave state in the single-band case, as a function of the momentum along the surface.}
\label{fig: ABS-energy}
\end{figure}

The calculations are particularly simple in the case of ${\cal N}=1$, when the scattering matrix becomes just a single complex number (a pure phase). The energy equation then has the form
\begin{equation}
\label{ABS-energy-eq}
  \frac{E+i\Omega}{E-i\Omega}=\frac{\Delta(\bk_{\mathrm{in}})}{\Delta(\bk_{\mathrm{out}})},
\end{equation}
which remarkably does not contain any surface scattering details. For a circular Fermi surface of radius $k_F$, we have 
$\bk_{\mathrm{in}}=k_F(-\cos\theta,\sin\theta)$, $\bk_{\mathrm{out}}=k_F(\cos\theta,\sin\theta)$, where $-\pi/2<\theta<\pi/2$ is the polar angle. Writing the gap function as $\Delta(\bk)=\Delta_0e^{iN\theta}$, where
$N$ is the phase winding number around the Fermi surface, we obtain from Eq. (\ref{ABS-energy-eq}) the following expression for the ABS energy:
\begin{equation}
\label{ABS-energy}
  E(\theta)=-\Delta_0\cos(N\theta)\sgn[\sin(N\theta)].
\end{equation}
For example, the chiral $d$-wave state, see Eq. (\ref{chiral-d-wave}), has $N=2$, and the ABS energy dispersion as a function of $k_y=k_F\sin\theta$ is shown in Fig. \ref{fig: ABS-energy}. The spectrum is discontinuous and has two nondegenerate branches,
passing through zero at $\theta=\pm\pi/4$, \textit{i.e.}, at $k_y=\pm k_F/\sqrt{2}$. According to Eq. (\ref{ABS-energy-eq}), the origin of the subgap surface states can be traced to the gap function having different values on the incident and 
reflected legs of the semiclassical trajectory, similar to the ABS in centrosymmetric $d$-wave superconductors\cite{Hu94} or near a superconducting domain wall.\cite{DW-fermions} The spectrum discontinuity occurs at $k_y=0$, 
where the ABS is absent due to the quasiparticles sensing the same gap function before and after the surface reflection.

In general, the ABS energy (\ref{ABS-energy}) vanishes at
$$
  \theta=\pm\frac{\pi}{2|N|}(2k+1),\quad k=0,...,\frac{|N|}{2}-1. 
$$
Therefore, if the phase winding number of the gap function in the bulk is equal to $N$, then there are exactly $|N|$ fermionic zero modes propagating in the same direction along the surface, 
which is a manifestation of the bulk-boundary correspondence in topological superconductors.\cite{Volovik-book} Note that $N$ is even, because the gap function is even in momentum.

\section{Bogoliubov Berry phase}
\label{sec: Berry phases}

Another type of topological invariants is obtained using the Berry connections of the Bogoliubov quasiparticle wave functions. 
Let us look at the Bogoliubov band labelled by the Bloch band index $n$ and the electron-hole index $s=\pm$. Assuming a nodeless gap function, the excitation energy is nonzero at all momenta. 
We introduce the Berry connection $1$-form, which is an analog of the magnetic vector potential:
\begin{equation}
\label{intraband Berry-A}
  i\langle\Phi_{n,s}(\bk)|d|\Phi_{n,s}(\bk)\rangle=A_n(\bk)+a_{n,s}(\bk).
\end{equation}
We used Eq. (\ref{Phi-ns}) to represent the last expression as a sum of the intraband Bloch Berry connection 
\begin{equation}
\label{A-n-def}
  A_n(\bk)=i\langle\bk,n|d|\bk,n\rangle={\cal A}_{nn}(\bk),
\end{equation}
see Eq. (\ref{Berry-A-form}), and the Nambu Berry connection 
\begin{equation}
\label{a-ns-def}
  a_{n,s}(\bk)=i\langle\bk,n;s|d|\bk,n;s\rangle.
\end{equation}
Both connections can be integrated along one-dimensional manifolds in momentum space. In our case, it is natural to integrate along the Fermi surfaces, which form closed loops on the BZ torus.  

One can define the Berry phase factor associated with the $(n,s)$th Bogoliubov band as the Wilson loop along the $n$th Fermi surface:
\begin{equation}
\label{Bogoliubov-Wilson}
  W_{n,s}=e^{i\oint_{FS_n}(A_n+a_{n,s})}.
\end{equation}
Each Fermi surface is assumed to be connected, with the orientation defined by $\hat z\times\bm{v}_{F,n}$.  
In the next two subsections we will prove that the Bogoliubov Wilson loop has the following form:
\begin{equation}
\label{Bogoliubov-Wilson-final}
  W_{n,s}=(-1)^{M_n+N_n},
\end{equation}
where $M_n$ is the number of the band degeneracies enclosed by the $n$th Fermi surface and $N_n$ is the phase winding number of the nodeless gap function $\Delta_n(\bk)$ around the Fermi surface.

\subsection{Bloch Berry phase}
\label{sec: Bloch Berry}

The Bloch Berry phase factor
\begin{equation}
\label{Berry-Wilson}
  w^B_n=e^{i\oint_{FS_n}A_n}
\end{equation}
is invariant under the $U(1)$ gauge transformations, in which the phase of the Bloch state $|\bk,n\rangle$ can change by an integer multiple of $2\pi$ as one goes around the Fermi surface. To calculate $w^B_n$, we introduce another 
gauge-invariant quantity, called the Berry curvature of the connection $A_n$ (Ref. \onlinecite{Berry-phase-book}). It is defined as $dA_n=(1/2)F_{ij}dk_idk_j$ (the exterior product sign is again omitted for brevity), 
where the field strength tensor has only two nonzero components: $F_{n,xy}=-F_{n,yx}=B_n$. Thus the Berry ``magnetic field'' is directed along the $z$ axis and given by
\begin{equation}
\label{Bloch-B}
  B_n(\bk)=i\left(\frac{\partial\langle\bk,n|}{\partial k_x}\frac{\partial|\bk,n\rangle}{\partial k_y}-\frac{\partial\langle\bk,n|}{\partial k_y}\frac{\partial|\bk,n\rangle}{\partial k_x}\right).
\end{equation}
For a nondegenerate band, the flux of the Berry field through the entire BZ torus is a topological invariant, called the first Chern number. It is also known as the TKNN integer, after Ref. \onlinecite{TKKN82}, where it was shown to   
be responsible for the quantization of Hall conductance in the presence of magnetic field.

In our case, the electron bands are invariant under TR, which means that $B_n(-\bk)=-B_n(\bk)$ (Ref. \onlinecite{Berry-phase-book}). On the other hand, the Bloch Berry field transforms as a vector under proper rotations. 
In particular, if the 2D point group of the system contains the twofold rotation $C_{2z}$, which acts as a ``2D inversion'': $C_{2z}(k_x,k_y)=(-k_x,-k_y)$, then $B_n(-\bk)=B_n(\bk)$. 
Therefore, for $G=\mathbf{C}_2$, $\mathbf{C}_4$, $\mathbf{C}_6$, $\mathbf{D}_2$, $\mathbf{D}_4$, and $\mathbf{D}_6$, we have
\begin{equation}
\label{Berry B zero}
  B_n(\bk)=0
\end{equation}
everywhere except the TR invariant points $\bK_i$, where the $n$th band is degenerate with another band, see Sec. \ref{sec: Rashba model}, and the Berry field is not defined. Now one can use Stokes' theorem to contract the integration contour 
in Eq. (\ref{Berry-Wilson}) by deforming it through the degeneracy-free regions of the BZ and show that
\begin{equation}
\label{fl-to-ci}
  \oint_{FS_n}A_n=\sum_i{}'\oint_{c_i}A_n.
\end{equation}
Here the prime means that the sum is taken only over the TR invariant points enclosed by the $n$th Fermi surface and $c_i$ is an infinitesimally small circular contour around $\bK_i$, whose orientation is inherited from the Fermi surface.
Therefore, the Berry phase is completely determined by the behavior of the Bloch wave functions in the vicinity of the band degeneracies.
Note that, according to Eq. (\ref{Berry B zero}), the Berry flux through the entire punctured BZ (with $\bK_i$'s removed) vanishes.  Then it follows from Stokes' theorem that
\begin{equation}
\label{sum-all-Ki}
  \sum_{i=1}^{N_{\mathrm{TRI}}}\oint_{c_i}A_n=0,
\end{equation}
where the summation is over all TR invariant points in the BZ.

It is easy to see that the above results also hold for the remaining 2D point groups, $G=\mathbf{C}_1$, $\mathbf{C}_3$, $\mathbf{D}_1$, and $\mathbf{D}_3$. In these cases, the Berry field is not identically zero at all $\bk\neq\bK_i$, but 
its fluxes through the punctured interior of the Fermi surface as well as through the entire punctured BZ still vanish due to the TR invariance. Since the integration contour in Eq. (\ref{Berry-Wilson}) can be deformed into $c_i$ in such a way that 
it stays TR invariant (\textit{i.e.}, symmetric with respect to $\bk\to-\bk$), the Berry flux through the contour remains zero and we again arrive at Eq. (\ref{fl-to-ci}). 

As an illustration, let us consider the generalized Rashba model, see Eq. (\ref{H-Rashba}), with just two bands labelled by $n=\lambda=\pm$. Using the spherical angle parametrization of the 
SO coupling and the eigenstates (\ref{Bloch-Rashba}), we obtain for the Berry connections:
\begin{equation}
\label{Bloch-A-Rashba}
  A_\lambda=-\frac{1}{2}(1-\lambda\cos\alpha)d\beta.
\end{equation}
For the 2D point groups containing the rotation $C_{2z}$, the antisymmetric SO coupling is planar, see Sec. \ref{sec: Rashba model}, \textit{i.e.}, $\gamma_z=0$ and $\alpha=\pi/2$, therefore, $A_\lambda=-d\beta/2$. 
Neither $\beta$ nor $A_\lambda$ are defined at the band degeneracies, but everywhere else the Berry connection form is locally exact, resulting in the Berry field $B_\lambda$ being zero in the entire punctured BZ. 
For the point groups not containing $C_{2z}$, the SO coupling is nonplanar and the Berry field is given by
$$
  B_\lambda(\bk)=-\frac{\lambda}{2}\sin\alpha\left(\frac{\partial\alpha}{\partial k_x}\frac{\partial\beta}{\partial k_y}-\frac{\partial\alpha}{\partial k_y}\frac{\partial\beta}{\partial k_x}\right)=-B_\lambda(-\bk).
$$
The last equality is due to TR invariance, which implies $\bgam(-\bk)=-\bgam(\bk)$, therefore $\alpha(-\bk)=\pi-\alpha(\bk)$ and $\beta(-\bk)=\pi+\beta(\bk)$. Then it follows from Eq. (\ref{Bloch-A-Rashba}) that 
\begin{equation}
\label{int-A-Rashba}
  \oint_{c_i}A_\lambda=-\pi N_\beta(\bK_i),
\end{equation}
where $N_\beta$ is the winding number of the planar component $\bgam_\parallel=(\gamma_x,\gamma_y,0)$ of the antisymmetric SO coupling around $\bK_i$. The contribution to the integral of the second term in $A_\lambda$ vanishes, because it is odd
under the inversion with respect to $\bK_i$: $\bgam(\bK_i-\bq)=\bgam(\bK_i+\bm{G}-\bq)=\bgam(-\bK_i-\bq)=-\bgam(\bK_i+\bq)$. 
Inserting the expression (\ref{int-A-Rashba}) in Eqs. (\ref{Berry-Wilson}) and (\ref{fl-to-ci}), we obtain for the Berry Wilson loop in the helicity bands:
\begin{equation}
\label{w-Rashba}
  w_\lambda^B=(-1)^{\sum_i'N_\beta(\bK_i)}.
\end{equation}
This is a $Z_2$ invariant determined by the total winding number of $\beta$ for the band degeneracies inside the Fermi surface. 

The winding number $N_\beta(\bK_i)$ is nothing but the index of the critical point $\bK_i$ of the vector field $\bgam_\parallel(\bk)$. 
In the vicinity of $\bK_i$, one can write $\gamma_{\parallel,l}(\bK_i+\bq)=\mu_{i,lm}q_m$, where $l,m=x,y$. Then,
\begin{equation}
\label{N-index}
  N_\beta(\bK_i)=\sgn\det\hat\mu_i, 
\end{equation}
see Ref. \onlinecite{DFN85}. We assume that $c_i$ is oriented counterclockwise and that the matrix $\hat\mu_i$ is nondegenerate, which is a generic situation, according to Table \ref{table gammas}. 
For example, from Eqs. (\ref{gamma-square}) and (\ref{gamma-hexagonal}) we obtain: 
$$
  N_\beta(\bK_1)=N_\beta(\bK_4)=1,\quad N_\beta(\bK_2)=N_\beta(\bK_3)=-1
$$
for the square lattice, and
$$
    N_\beta(\bK_1)=N_\beta(\bK_5)=N_\beta(\bK_6)=1,\quad N_\beta(\bK_2)=N_\beta(\bK_3)=N_\beta(\bK_4)=-1
$$
for the hexagonal lattice. In both cases, the sum of the winding numbers for all $\bK_i$ vanishes, in agreement with Eq. (\ref{sum-all-Ki}). This is just a consequence 
of the Poincare theorem,\cite{DFN85} according to which the sum of the indices for all critical points is equal to the Euler characteristic of the BZ, which is zero. 

It follows from Eq. (\ref{N-index}) that $N_\beta=\pm 1$ and, therefore, the $Z_2$ invariant (\ref{w-Rashba}) is equal to $+1$ ($-1$), if the Fermi surface encloses an even (odd) number of the TR invariant points. 
Remarkably, this result also holds in the general case of arbitrary bands in a noncentrosymmetric crystal, beyond the model (\ref{H-Rashba}). Indeed, in the vicinity of an isolated point $\bK_i$, where two Bloch bands cross, 
one can neglect all other bands and describe the remaining two-band structure using a generalized Rashba model, as illustrated in Appendix \ref{sec: 4-band model}. 
This allows one to represent the Berry connection near $\bK_i$ in the form (\ref{Bloch-A-Rashba}), which means that its loop integral is quantized as in Eq. (\ref{int-A-Rashba}):
$$
  \oint_{c_i}A_n=\pm\pi.
$$
Therefore,
\begin{equation}
\label{w-n-final}
  w^B_n=(-1)^{M_n},
\end{equation}
where $M_n$ is the number of the TR invariant points enclosed by the nondegenerate Fermi surface in the $n$th band.

One can also define a $Z_2$ invariant for the whole BZ, see Appendix \ref{sec: Z2-bands}, which does not depend on the position of the Fermi level and can therefore be used for a topological 
classification of the Bloch bands in any crystal, whether superconducting, metallic, or insulating.
In contrast to the Wilson loop (\ref{w-n-final}), the whole-BZ invariant is determined by the parity of the number of the TR invariant points inside the HBZ.

\subsection{Nambu Berry phase}
\label{sec: Nambu Berry}

The Nambu Berry phase is given by the Wilson loop
\begin{equation}
\label{Nambu-Wilson}
  w^N_{n,s}=e^{i\oint_{FS_n}a_{n,s}},
\end{equation}
which is invariant under the $U(1)$ gauge transformation of the Nambu eigenstates $|\bk,n;s\rangle$.
Introducing two spherical angles to parametrize Eq. (\ref{h-nu}): $\hat{\bm{\nu}}_n=(\sin\tilde\alpha_n\cos\tilde\beta_n,\sin\tilde\alpha_n\sin\tilde\beta_n,\cos\tilde\alpha_n)$, we obtain the following expression for the Nambu connection:
\begin{equation}
\label{Nambu-a}
  a_{n,s}=-\frac{1}{2}(1-s\cos\tilde\alpha_n)d\tilde\beta_n,
\end{equation}
cf. Eq. (\ref{Bloch-A-Rashba}). At the Fermi surface, $\xi_n=0$ and the connection is exact: $a_{n,s}=-d\tilde\beta_n/2$. For a nodeless gap function given by Eq. (\ref{nodeless gap}), we have $\tilde\beta_n=-\varphi_n$ and 
$$
  \oint_{FS_n}a_{n,s}=\pi N_n,
$$
where $N_n$ is the phase winding number, see Eq. (\ref{winding-number}). Therefore, 
\begin{equation}
\label{Nambu-Wilson-final}
  w^N_{n,s}=(-1)^{N_n}.
\end{equation}
Now one can calculate the total Berry phase factor of the Bogoliubov quasiparticles in the $n$th band, $W_{n,s}=w^B_nw^N_{n,s}$. Combining Eqs. (\ref{w-n-final}) and (\ref{Nambu-Wilson-final}), we arrive at the expression (\ref{Bogoliubov-Wilson-final}).

The case of a TR invariant superconducting state, corresponding to real gap functions $\Delta_n$, requires a different treatment, because the phase winding numbers all vanish and $w^N_{n,s}=1$. Instead of the Wilson loop (\ref{Nambu-Wilson}), 
one can introduce another natural invariant -- the sign of the gap in the $n$th band, which cannot be changed without the gap closing. Then, Eq. (\ref{Bogoliubov-Wilson-final}) is replaced by 
$W^{\mathrm{TRI}}_n=(-1)^{M_n}\sgn(\Delta_n)$, so that
\begin{equation}
   W_{\mathrm{TRI}}=\prod_n(-1)^{M_n}\sgn(\Delta_n)
\end{equation}
is a $Z_2$ invariant. In particular, in the two-band case, assuming $M_+=M_-=1$, we have $W_{\mathrm{TRI}}=\sgn(\Delta_+\Delta_-)$. Therefore, the superconducting state is topologically nontrivial ($W_{\mathrm{TRI}}=-1$) 
if the two gap functions have opposite signs. This result is consistent with Refs. \onlinecite{SF09,TYBN09}, and \onlinecite{QHZ10}, where a $Z_2$ invariant in the TR invariant case was derived following different routes.

\section{Conclusions}
\label{sec: conclusions}

We developed a theory of superconductivity in 2D metals lacking the mirror symmetry under the reflection $z\to -z$. Strong SO coupling of electrons with the crystal lattice results in the splitting of the Bloch bands, 
which is much larger than the energy scales associated with superconductivity. Spin or pseudospin is no longer a good quantum number, and the Cooper pairing occurs only between the time-reversed states of the same helicity. 
Constructing the pairing Hamiltonian and the gap functions from the exact Bloch band states not only naturally fits into the framework of the Fermi-liquid and the BCS theories, but also highlights the novel features of noncentrosymmetric superconductors.

Superconducting states are classified according to the irreducible representations of the 2D point groups, but the results differ significantly from the textbook case of centrosymmetric superconductors. 
In particular, the gap functions are always even in momentum, and not every irreducible representation corresponds to a possible pairing symmetry.
We also derived the proper BdG Hamiltonian for noncentrosymmetric superconductors in the band representation, avoiding any ambiguity about the choice of the gap functions. 

We studied the momentum-space topology of 2D noncentrosymmetric superconductors using two different topological invariants. The first one is 
the Maurer-Cartan invariant built from the BdG Green's functions. It takes integer values, determined by the winding numbers of the gap function phases around the Fermi surfaces, and is independent of the Berry curvature 
of the Bloch states. The absolute value of the phase winding number is equal to the number of fermionic subgap boundary modes propagating in the same direction along the surface of the superconductor. We calculated the spectrum of these modes in the 
semiclassical approximation, solving the Andreev equations in the helical bands with the boundary conditions expressed in terms of the surface scattering matrix. The boundary modes can provide an experimentally verifiable 
signature of a topologically nontrivial superconducting state.  
The second type of the topological invariants is given by the Wilson loops of the Berry vector potential of the BdG eigenstates. In each band, the Wilson loop is equal to the parity of 
the sum of the number of the band degeneracies (Weyl points) enclosed by the Fermi surface and the gap phase winding number. The helicity bands always remain pairwise degenerate at the TR invariant points, the latter serving as the sources 
of a topological ``twist'' in the band wave functions, whose positions in the BZ are fixed by the crystal symmetry. 

\acknowledgments

This work was supported by a Discovery Grant from the Natural Sciences and Engineering Research Council of Canada.

\appendix

\section{Four-band model}
\label{sec: 4-band model}

We begin with two bands, $\epsilon_1(\bk)$ and $\epsilon_2(\bk)$, satisfying $\epsilon_1(\bk)<\epsilon_2(\bk)$, each twofold degenerate due to pseudospin, and turn on the inversion-antisymmetric part of the lattice potential 
along with the corresponding SO coupling. In Eq. (\ref{H pseudospin}), we keep only $\mu,\nu=1,2$ and introduce the following notations:
\begin{eqnarray*}
  && A_{11}=A_{22}=0,\quad A_{12}=-A_{21}=\ell,\\
  && \bm{B}_{11}=\bgam_1,\quad \bm{B}_{22}=\bgam_2,\quad \bm{B}_{12}=\bm{B}_{21}=\tilde{\bgam}.
\end{eqnarray*}
Here the scalar $\ell(\bk)$ and the pseudovectors $\bgam_{1,2}(\bk),\tilde{\bgam}(\bk)$ are real, odd in $\bk$, and invariant under the point group operations. 
The Hamiltonian (\ref{H pseudospin}) takes the form of two coupled Rashba models: 
\begin{equation}
\label{H-4-band}
  \hat H_0=\hat H_1+\hat H_2+\hat H_{12},
\end{equation}
where
$$
  \hat H_\mu=\sum\limits_{\bk,\alpha\beta}[\epsilon_\mu(\bk)\delta_{\alpha\beta}+\bgam_\mu(\bk)\bm{\sigma}_{\alpha\beta}]\hat a^\dagger_{\bk\mu\alpha}\hat a_{\bk\mu\beta}
$$
and 
$$
  \hat H_{12}=\sum\limits_{\bk,\alpha\beta}[i\ell(\bk)\delta_{\alpha\beta}+\tilde{\bgam}(\bk)\bm{\sigma}_{\alpha\beta}]\hat a^\dagger_{\bk 1\alpha}\hat a_{\bk 2\beta}+\mathrm{H.c.}
$$
Representative expressions for the pseudovector SO couplings near the $\Gamma$ point can be found in Table \ref{table gammas}, while those for the scalar coupling are given in Table \ref{table ells}. 
The latter vanishes identically for the point groups containing the rotation $C_{2z}$, because $\ell(\bk)=\ell(C_{2z}^{-1}\bk)=\ell(-\bk)=-\ell(\bk)$, therefore $\ell(\bk)=0$. 
Diagonalizing Eq. (\ref{H-4-band}), we will obtain four bands, which remain twofold degenerate at the TR invariant points, where $\ell,\bgam_{1,2}$, and $\tilde{\bgam}$ all vanish.  

One can make analytical progress by assuming that, due to their symmetry properties being the same, all three pseudovector SO couplings have the same 
momentum dependence: $\bgam_{1,2}(\bk)=\gamma_{1,2}\mathbf{g}(\bk)$ and $\tilde{\bgam}(\bk)=\tilde\gamma\mathbf{g}(\bk)$, where $\gamma_{1,2}$ and $\tilde\gamma$ are positive constants.
Then we obtain the following band dispersions:
\begin{equation}
\label{4-band-dispersions}
  \xi_{1,2}(\bk)=r_\mp(\bk)-s_\mp(\bk),\quad \xi_{3,4}(\bk)=r_\mp(\bk)+s_\mp(\bk), 
\end{equation}
where 
$$
  r_\pm=\frac{\epsilon_1+\epsilon_2}{2}\pm\frac{\gamma_1+\gamma_2}{2}|\mathbf{g}|
$$
and
$$
  s_\pm=\sqrt{\left(\frac{\epsilon_1-\epsilon_2}{2}\pm\frac{\gamma_1-\gamma_2}{2}|\mathbf{g}|\right)^2+\tilde\gamma^2|\mathbf{g}|^2+\ell^2}.
$$
Near the TR invariant points, the band structure is given in the leading approximation by $\xi_{1,2}(\bk)=\epsilon_1(\bk)\pm\gamma_1|\mathbf{g}|$ and $\xi_{3,4}(\bk)=\epsilon_2(\bk)\pm\gamma_2|\mathbf{g}|$, 
with $\xi_1(\bk)\leq\xi_2(\bk)<\xi_3(\bk)\leq\xi_4(\bk)$,
\textit{i.e.}, the spectrum is entirely determined by the intraband antisymmetric SO couplings of the Rashba type.

\begin{table}
\caption{The scalar antisymmetric interband coupling near the center of the BZ ($a_{1,2}$ and $a$ are real
constants, $b$ is a complex constant, and $k_\pm=k_x\pm ik_y$). }
\begin{tabular}{|c|c|}
    \hline
    \hspace*{2mm} $G$\hspace*{2mm}    & $\ell(\bk)$  \\ \hline
    $\mathbf{C}_1$      & $a_1k_x+a_2k_y$ \\ \hline
    $\mathbf{C}_3$    & $bk_+^3+b^*k_-^3$  \\ \hline
    $\mathbf{D}_1$      & $ak_y$  \\ \hline
    $\mathbf{D}_3$     & \hspace*{1mm} $ia(k_+^3-k_-^3)$ \hspace*{1mm} \\ \hline
    \hspace*{1mm} $\mathbf{C}_{2,4,6}$, $\mathbf{D}_{2,4,6}$ \hspace*{1mm} & $0$ \\ \hline
\end{tabular}
\label{table ells}
\end{table}

\section{Maurer-Cartan topological invariants}
\label{sec: MC-inv}

Let us consider a map of a $D$-dimensional manifold ${\cal M}$, supplied with coordinates $\bm{x}=(x^1,x^2,...,x^D)$, 
into a manifold of invertible square matrices $\hat g(\bm{x})$. We also assume that there exists a constant matrix $\hat S$, which either commutes or anticommutes with $\hat g$ at all $\bm{x}$. 
The matrix $\hat S$ is then represents a ``symmetry'' of $\hat g$. The matrix-valued Maurer-Cartan 1-form is defined by 
$\hat\omega\equiv\hat g^{-1}d\hat g$. One can then introduce the following scalar-valued $D$-forms on ${\cal M}$:
\begin{equation}
\label{MC-D-form}
  \Omega_D=\tr\hat S\hat\omega^D,
\end{equation}
where the matrix trace is taken and the powers of $\hat\omega$ should be understood in the sense of combined exterior and matrix multiplication. 
Using cyclic invariance of the trace in Eq. (\ref{MC-D-form}) and antisymmetry of the exterior product, it is easy to show that $\Omega_D=(-1)^{D-1}\Omega_D$, therefore $\Omega_D=0$ for even $D$.

In odd dimensions, we integrate Eq. (\ref{MC-D-form}) over ${\cal M}$ to obtain:
\begin{equation}
\label{MC-I-D}
  I_D[\hat g]=\int_{{\cal M}}\Omega_D.
\end{equation}
Now let us show that if ${\cal M}$ is a closed manifold, then the above expression is invariant under small variations 
$\hat g\to\hat g+\delta\hat g$ which respect the symmetry of $\hat g$, \textit{i.e.}, $\delta\hat g$ also either commutes or anticommutes with $\hat S$. Keeping only the terms linear in $\delta\hat g$, we find 
$\delta\hat\omega=d\hat\rho+[\hat\omega,\hat\rho]$, where $\hat\rho=\hat g^{-1}\delta\hat g$ is a matrix ($0$-form), and 
\begin{equation}
\label{delta Omega-D}
  \delta\Omega_D=D\tr\hat S\delta\hat\omega\hat\omega^{D-1}=D\tr\hat Sd\hat\rho\hat\omega^{D-1}+D\tr\hat S[\hat\omega,\hat\rho]\hat\omega^{D-1}.
\end{equation}
Since $\hat\omega$ commutes with $\hat S$ and $D$ is odd, the second term on the right-hand side vanishes. Taking into account the product rule of differentiation for differential forms, 
$d(ab)=(da)b+(-1)^{\mathrm{deg}\,a}a(db)$, we obtain from Eq. (\ref{delta Omega-D}):
$$
  \delta\Omega_D=D\tr\hat S d(\hat\rho\hat\omega^{D-1})-D\tr\hat S\hat\rho d\hat\omega^{D-1}.
$$
One can use the product rule again, along with the property $d\hat\omega=-\hat\omega^2$, to show that the second term here is zero. Indeed,
\begin{eqnarray*}
  d\hat\omega^{D-1}=d\hat\omega\hat\omega^{D-2}-\hat\omega d\hat\omega^{D-2}=-\hat\omega^D-\hat\omega(d\hat\omega\hat\omega^{D-3}-\hat\omega d\hat\omega^{D-3})=\hat\omega^2d\hat\omega^{D-3}=...=0,
\end{eqnarray*}
because $D-1$ is even. Thus we have shown that $\delta\Omega_D$ is exact:
$\delta\Omega_D=d\tilde\Omega_{D-1}$, where $\tilde\Omega_{D-1}=D\tr\hat S\hat\rho\hat\omega^{D-1}$. If ${\cal M}$ is closed, then, according to Stokes' theorem,
$$
  \delta I_D=\int_{{\cal M}}\delta\Omega_D=\int_{\partial{\cal M}}\tilde\Omega_{D-1}=0.
$$
Therefore, the expression (\ref{MC-I-D}) is a topological invariant, said to be ``protected'' by the symmetry $\hat S$. If $\hat g$ does not have any symmetries, then $\hat S$ in the above expressions is just the unit matrix,
which is what we assumed in Sec. \ref{sec: GFs}.

\section{Sketch of the derivation of Eq. (\ref{Tr-omega3-K})}
\label{sec: long derivation}

It follows from the representation (\ref{omega-PQ}) of the Maurer-Cartan form that
\begin{eqnarray}
  \Tr\omega^3 &=& \sum_{m_{1,2,3}n_{1,2,3}}\trb(\hat\Pi_{m_1}d\hat\Pi_{n_1}\hat\Pi_{m_2}d\hat\Pi_{n_2}\hat\Pi_{m_3}d\hat\Pi_{n_3})\trn(\hat P_{m_1n_1}\hat P_{m_2n_2}\hat P_{m_3n_3})\nonumber\\
      && +\sum_{m_{1,2}n_{1,2,3}}\trb(\hat\Pi_{m_1}d\hat\Pi_{n_1}\hat\Pi_{m_2}d\hat\Pi_{n_2}\hat\Pi_{n_3})\trn(\hat P_{m_1n_1}\hat P_{m_2n_2}\hat Q_{n_3})\nonumber\\
      && -\sum_{m_{1,3}n_{1,2,3}}\trb(\hat\Pi_{m_1}d\hat\Pi_{n_1}\hat\Pi_{n_2}\hat\Pi_{m_3}d\hat\Pi_{n_3})\trn(\hat P_{m_1n_1}\hat Q_{n_2}\hat P_{m_3n_3})\nonumber\\
      && +\sum_{m_{1}n_{1,2,3}}\trb(\hat\Pi_{m_1}d\hat\Pi_{n_1}\hat\Pi_{n_2}\hat\Pi_{n_3})\trn(\hat P_{m_1n_1}\hat Q_{n_2}\hat Q_{n_3})\nonumber\\
      && +\sum_{m_{2,3}n_{1,2,3}}\trb(\hat\Pi_{n_1}\hat\Pi_{m_2}d\hat\Pi_{n_2}\hat\Pi_{m_3}d\hat\Pi_{n_3})\trn(\hat Q_{n_1}\hat P_{m_2n_2}\hat P_{m_3n_3})\nonumber\\
      && -\sum_{m_{2}n_{1,2,3}}\trb(\hat\Pi_{n_1}\hat\Pi_{m_2}d\hat\Pi_{n_2}\hat\Pi_{m_3})\trn(\hat Q_{n_1}\hat P_{m_2n_2}\hat Q_{n_3})\nonumber\\
      && +\sum_{m_{3}n_{1,2,3}}\trb(\hat\Pi_{n_1}\hat\Pi_{n_2}\hat\Pi_{m_3}d\hat\Pi_{n_3})\trn(\hat Q_{n_1}\hat Q_{n_2}\hat P_{m_3n_3})\nonumber\\
      && +\sum_{n_{1,2,3}}\trb(\hat\Pi_{n_1}\hat\Pi_{n_2}\hat\Pi_{n_3})\trn(\hat Q_{n_1}\hat Q_{n_2}\hat Q_{n_3}).\label{Tr-omega-Pis}
\end{eqnarray}
Here ``$\trb$'' and ``$\trn$'' denote the matrix traces in the Bloch ($L_B$) and Nambu ($L_N$) spaces, respectively. 
The sign change in the third and sixth terms is due to the anticommutation of 1-forms. The Bloch traces on the right-hand side can be calculated using the Berry connection, see Eq. (\ref{Berry-A-form}), and the following properties 
of the band projectors:
$$
  \sum_n\hat\Pi_n=\mathbb{1}_B,\quad \hat\Pi_m\hat\Pi_n=\delta_{mn}\hat\Pi_n,\quad \hat\Pi_m d\hat\Pi_n \hat\Pi_m=0,
$$
where $\mathbb{1}_B$ is the unit operator in $L_B$. For instance, we have
\begin{eqnarray*}
  &&\trb(\hat\Pi_{m_1}d\hat\Pi_{n_1}\hat\Pi_{m_2}d\hat\Pi_{n_2}\hat\Pi_{m_3}d\hat\Pi_{n_3})=(\delta_{n_1m_2}{\cal A}_{m_1n_1}-\delta_{m_1n_1}{\cal A}_{n_1m_2})\\
  && \hspace*{3cm}\times(\delta_{n_2m_3}{\cal A}_{m_2n_2}-\delta_{m_2n_2}{\cal A}_{n_2m_3})(\delta_{n_3m_1}{\cal A}_{m_3n_3}-\delta_{m_3n_3}{\cal A}_{n_3m_1}),\\
  &&\trb(\hat\Pi_{m_1}d\hat\Pi_{n_1}\hat\Pi_{n_2}\hat\Pi_{n_3})=0,
\end{eqnarray*}
\textit{etc}. Collecting similar terms in Eq. (\ref{Tr-omega-Pis}), we obtain:
\begin{equation}
\label{K-023}
  \Tr\omega^3=\sum_nK^{(0)}_n+\sum_{mn}K^{(2)}_{mn}{\cal A}_{mn}{\cal A}_{nm}+\sum_{lmn}K^{(3)}_{lmn}{\cal A}_{lm}{\cal A}_{mn}{\cal A}_{nl},
\end{equation}
where
\begin{eqnarray*}
  && K^{(0)}_n = \trn \hat Q_n^3,\\
  && K^{(2)}_{mn} =-\trn\bigl(2\hat P_{mn}\hat P_{nm}\hat Q_m-2\hat P_{mn}\hat Q_m-2\hat P_{nm}\hat Q_m+2\hat Q_m-\hat P_{mn}\hat Q_n\hat P_{nm}+\hat P_{mn}\hat Q_n+\hat Q_n\hat P_{nm}-\hat Q_n\bigr),\\
  && K^{(3)}_{lmn} = i\trn\bigl(\hat P_{lm}\hat P_{mn}\hat P_{nl}-\hat P_{lm}\hat P_{mn}-\hat P_{mn}\hat P_{nl}-\hat P_{nl}\hat P_{lm}+\hat P_{lm}+\hat P_{mn}+\hat P_{nl}\bigr)-2i.
\end{eqnarray*}
Inserting here Eq. (\ref{PQ}), calculating the Nambu traces, and using the fact that, since the Berry connection 1-forms anticommute, only the part of $K^{(2)}_{mn}$ which is antisymmetric under the interchange $m\leftrightarrow n$ 
enters the right-hand side of Eq. (\ref{K-023}), we finally arrive at the expression (\ref{Tr-omega3-K}).

\section{$Z_2$ invariant for the whole BZ}
\label{sec: Z2-bands}

We focus on the $n$th Bloch band in a 2D noncentrosymmetric crystal. According to Eq. (\ref{t_n}), the Bloch states $K|\bk,n\rangle$ and $|-\bk,n\rangle$ differ by a phase factor $t_n(\bk)$, which is not 
defined at the TR invariant points $\bK_i$, where $i=1,...,N_{\mathrm{TRI}}$. We want to find whether the maps $\bk\to t_n(\bk)$ can fall into different topological classes. 
Since $t_n(-\bk)=-t_n(\bk)$, it is sufficient to consider just the maps of the HBZ.

We introduce the 1-form $\alpha_n=-id\ln t_n$, which can be integrated along the HBZ boundary to define
$$
  D_n=-\frac{1}{2\pi}\biggl[\oint_{\partial(\mathrm{HBZ})}\alpha_n\biggr]\;\mathrm{mod}\ 2.
$$
Any singularities in the integrand that are due to the TR invariant points located at the HBZ boundary can be removed by redefining (shifting) the HBZ. 
However, regardless of how one redefines the HBZ, half of the TR invariant points will always remain in its interior. Removing infinitesimal vicinities of these points, the form $\alpha_n$ becomes locally exact in the punctured HBZ
and one obtains from Stokes' theorem:
\begin{equation}
\label{alpha-winding}
  D_n=-\frac{1}{2\pi}\Biggl(\sum_{i=1}^{N_{\mathrm{TRI}}/2}\oint_{c_i}\alpha_n\Biggr)\;\mathrm{mod}\ 2.
\end{equation}
The sum here is taken over the band degeneracies inside the HBZ. Under a gauge transformation of the Bloch states, $|\bk,n\rangle\to e^{i\theta_n(\bk)}|\bk,n\rangle$, 
where $e^{i\theta_n}$ is a single-valued function in the punctured BZ, we have $\alpha_n(\bk)\to\alpha_n(\bk)-d[\theta_n(\bk)+\theta_n(-\bk)]$, which leaves $D_n$ unchanged.

As discussed in Sec. \ref{sec: Bloch Berry}, one can describe the band structure in the vicinity of any band degeneracy point using an effective generalized
Rashba model, in which the planar component of the antisymmetric SO coupling is parametrized by an angle $\beta(\bk)$. Then it follows from Eqs. (\ref{t-factor-Rashba}) and (\ref{alpha-winding}) that $\alpha_n=-d\beta$ and
\begin{equation}
\label{alpha-winding-final}
  D_n=\Biggl[\sum_{i=1}^{N_{\mathrm{TRI}}/2}N_\beta(\bK_i)\Biggr]\;\mathrm{mod}\ 2.
\end{equation}
Since the winding number of $\beta$ around a generic band degeneracy is either $+1$ or $-1$, we obtain that $D_n=0$ if the number of the band degeneracies in the HBZ is even, and $D_n=1$ if this number is odd.

Another way of calculating the same $Z_2$ invariant involves integrals of the Bloch Berry connection and curvature, see Eqs. (\ref{A-n-def}) and (\ref{Bloch-B}), and is based on the following quantity:
\begin{equation}
\label{Z2-FK}
  \tilde D_n=\frac{1}{\pi}\biggl[\int_{\mathrm{HBZ}}B_n-\oint_{\partial(\mathrm{HBZ})}A_n\biggr]\;\mathrm{mod}\ 2,
\end{equation}
which characterizes an ``obstruction'' to the Stokes' theorem over the HBZ.\cite{FK06,top-SC}
This last expression makes sense only if one can find a way of dealing with the TR invariant points $\bK_i$ at the 
HBZ boundary and in its interior, as neither $A_n$ nor $B_n$ are defined at those points. We shift the HBZ as explained above, to make sure that its boundary avoids any band degeneracies, and perform the integration in the 
first term over the punctured HBZ. Then, Eq. (\ref{Z2-FK}) becomes
\begin{equation}
  \tilde D_n=-\frac{1}{\pi}\Biggl(\sum_{i=1}^{N_{\mathrm{TRI}}/2}\oint_{c_i}A_n\Biggr)\;\mathrm{mod}\ 2=\Biggl[\sum_{i=1}^{N_{\mathrm{TRI}}/2} N_\beta(\bK_i)\Biggr]\;\mathrm{mod}\ 2=D_n,
\end{equation}
where we used the expression (\ref{int-A-Rashba}) for the Bloch Berry phase integral in the vicinity of a band degeneracy. 

Thus, we see that the Bloch band is topologically trivial ($D_n=0$) if $N_{\mathrm{TRI}}/2$ is even, and topologically nontrivial ($D_n=1$) if $N_{\mathrm{TRI}}/2$ is odd. According to Eqs. (\ref{TRI-points-square}) and (\ref{TRI-points-hexagonal}),
the former possibility is realized, for instance, in a square lattice, while the latter -- in a hexagonal lattice.

\end{document}